\documentclass[twocolumn]{aastex63}

\received{June 1, 2019}
\revised{January 10, 2019}
\accepted{\today}
\submitjournal{ApJ}

\shorttitle{Chemical evolution in MaNGA galaxies}
\shortauthors{Camps-Fariña et al.}
\graphicspath{{./}{figures/}}

\begin{document}

\title{Chemical evolution history of MaNGA galaxies}

\correspondingauthor{Artemi Camps-Fariña}
\email{arcamps@ucm.es}

\author{Artemi Camps-Fariña}
\affiliation{Instituto de Astronom\'ia, Universidad Nacional Aut\'onoma de M\'exico, Apartado Postal 70-264, CP 04510 Ciudad de M\'exico, M\'exico\\}
\affiliation{Departamento de F\'{i}sica de la Tierra y Astrof\'{i}sica, Universidad Complutense de Madrid, 28040 Madrid, Spain}

\author{Sebastián F. S\'{a}nchez}
\affiliation{Instituto de Astronom\'ia, Universidad Nacional Aut\'onoma de M\'exico, Apartado Postal 70-264, CP 04510 Ciudad de M\'exico, M\'exico\\}

\author{Alfredo Mej\'{i}a-Narv\'{a}ez}
\affiliation{Instituto de Astronom\'ia, Universidad Nacional Aut\'onoma de M\'exico, Apartado Postal 70-264, CP 04510 Ciudad de M\'exico, M\'exico\\}

\author{Eduardo A. D. Lacerda}
\affiliation{Instituto de Astronom\'ia, Universidad Nacional Aut\'onoma de M\'exico, Apartado Postal 70-264, CP 04510 Ciudad de M\'exico, M\'exico\\}

\author{Leticia Carigi}
\affiliation{Instituto de Astronom\'ia, Universidad Nacional Aut\'onoma de M\'exico, Apartado Postal 70-264, CP 04510 Ciudad de M\'exico, M\'exico\\}

\author{Gustavo Bruzual}
\affiliation{Instituto de Radioastronom\'ia y Astrof\'iısica, IRyA, UNAM, Morelia, M\'exico}

\author{Paola Alvarez-Hurtado}
\affiliation{Instituto de Astronom\'ia, Universidad Nacional Aut\'onoma de M\'exico, Apartado Postal 70-264, CP 04510 Ciudad de M\'exico, M\'exico\\}

\author{Niv Drory}
\affiliation{McDonald Observatory, The University of Texas at Austin, 1 University Station, Austin, TX 78712, USA}

\author{Richard R. Lane}
\affiliation{Centro de Investigaci\'on en Astronom\'ia, Universidad Bernardo O'Higgins, Avenida Viel 1497, Santiago, Chile}

 \author{Nicholas Fraser Boardman}
 \affiliation{Department of Physics \& Astronomy, University of Utah, Salt Lake City, UT 84112, USA}
 
\author{Guillermo A. Blanc}
 \affiliation{The Observatories of the Carnegie Institution for Science, 813 Santa
Barbara Street, Pasadena, CA 91101, USA}
 \affiliation{Departamento de Astronom\'ia, Universidad de Chile, Casilla 36-D, Santiago, Chile\\}

\begin{abstract}
We show the results of a study using the spectral synthesis technique study for the full MaNGA sample showing their Chemical Enrichment History (ChEH) as well as the evolution of the stellar mass-metallicity relation (MZR) over cosmic time.
We find that the more massive galaxies became enriched first and the lower mass galaxies did so later, producing a change in the MZR which becomes shallower in time. Separating the sample into morphology and star-forming status bins some particularly interesting results appear: The mass dependency of the MZR becomes less relevant for later morphological types, to the extent that it inverts for Sd/Irr galaxies, suggesting that morphology is at least as important a factor as mass in chemical evolution. The MZR for the full sample shows a flattening at the high-mass end and another at the low-mass range, but the former only appears for retired galaxies while the latter only appears for star-forming galaxies. 
We also find that the average metallicity gradient is currently negative for all mass bins but for low mass galaxies it was inverted at some point in the past, before which all galaxies had a positive gradient.
We also compare how diverse the ChEHs are in the different bins considered as well as what primarily drives the diversity: How much galaxies become enriched or how quickly they do so.

\end{abstract}

%% Keywords should appear after the \end{abstract} command. 
%% See the online documentation for the full list of available subject
%% keywords and the rules for their use.

%%%%%%%%%%%%%%%%% BODY OF PAPER %%%%%%%%%%%%%%%%%%

\section{Introduction} \label{sec:intro}
Galaxy evolution remains one of the key problems to be solved in astrophysics. It involves many physical processes which take place at diverse scales, from star formation feedback to interaction between galaxies. The composition of each galaxy is also very important: the total mass of stars, their ages as well as the amount of gas and how enriched by metals it is, all have an impact on the evolution of their host galaxy.

The chemical content of a galaxy offers a unique view into how the evolutionary processes of galaxies work. Practically all elements heavier than hydrogen and helium are produced as the end result of star formation and evolution \citep[eg.][]{Kobayashi2020}. The quantity of metals present in a galaxy and its distribution are tied to star formation episodes and more fundamentally to stellar evolution \citep[][and references therein]{Yates2013}. Other processes, such as inflows, outflows, and mixing \citep{Tremonti2004,Lilly2013, Minchev2013,Minchev2014, Maiolino2019} can also influence the values of the metallicity.
Disentangling the history of the chemical enrichment in a galaxy is therefore a great way to probe into the history of galaxy evolution.

Many studies regarding the metallicity in galaxies focus on the nebular metallicity, that is, measuring the metal content of ionized gas using emission line ratios and following known calibrations \citep{Tremonti2004, Mannucci2010, Sanchez2013, Dayal2013, Barrera-Ballesteros2017}. The other way to measure metallicity is to analyze the spectra of the stellar population to derive the metallicity of the ISM that they were formed in, as we do in this article \citep[][]{Gallazzi2005, Panter2007, GonzalezDelgado2014}. These two methods are complementary, the nebular metallicity offers a snapshot of the current state of the cumulative chemical enrichment in a galaxy and has the advantage of being easier to measure as a result of relying on emission lines, though this also has some drawbacks \citep{Kewley2008,Lopez-Sanchez2012, Blanc2015, Maiolino2019}.

Stellar metallicity, on the other hand, requires a more detailed analysis to fit the spectra to templates to derive the metallicity values. However, it has the unique capacity to trace the metallicity at earlier cosmological times. The reason for this is that the metallicity of the ISM at the time of birth for a star is effectively "locked-in" in the star, allowing us to associate stars' metallicity to their age by tracing the varying values for different ages, and therefore different look-back time (LBT). Stars continuously produce metals throughout their lifetime but these remain at their central region for the majority of their lifetime while the atmospheric layers dominate the emission.
Thanks to these phenomena, we can measure the chemical enrichment over time for each galaxy in our sample.

In order to study the chemical enrichment of galaxies using nebular metallicity, we need to measure galaxies at different redshifts to infer a ChEH \citep{Lu2015,Cullen2019,Pharo2019,Urrutia2019}. This has the advantage of not being tied to model-dependent fits, as is the case for stellar metallicity. However, it has the disadvantage of not tracing the evolution of the same group of galaxies, which makes it vulnerable to selection biases. One example is how the requirement for strong emission lines means that galaxies with strong star formation are generally selected. Additionally, direct method nebular metallicities rely on the observation of weak auroral emission lines to be precise which are difficult to observe at high redshifts. Strong line based determinations of the nebular metallicity rely on empirical calibrators which are not guaranteed to be valid at all cosmic epochs and which have uncertainties regarding the absolute value of the metallicity \citep[e.g.,][]{Kennicutt2003, Perez-Montero2005, Kewley2008, Bresolin2009}.

In this article, we present results for the evolution of the chemical enrichment of galaxies from the MaNGA survey \citep{Bundy2015}, a large sample of $\sim$10,000 galaxies, using the spectral synthesis technique. We study both the ChEH and the stellar MZR, which offer complementary insights into the evolution of the galaxies. The structure of this article is as follows:
Section \ref{sec:sample} describes the characteristics of the sample, the criteria used to refine it, and a brief description of the data. In Section \ref{sec:analysis} we explain the methodology applied to reduce and analyze the data. In Section \ref{sec:results} we show the results, the estimated ChEH (\ref{sec:cheh}) and MZR (\ref{sec:mzr}) with sections \ref{sec:morph_results} and \ref{sec:sfs} showing how these parameters change if the divide the sample depending on morphology or star-forming status. We also show how measuring the metallicity at different galactocentric radii affects the results (Section \ref{sec:radius}) and how similar the chemical enrichment histories are within the bins that we consider for the sample in terms of mass, morphology and star-forming status (Section \ref{sec:variance}). Finally, in Section \ref{sec:discussion} we discuss the implications of the results and in Section \ref{sec:conclusions} we summarize our results.

\section{Sample and Data} \label{sec:sample}
The MaNGA survey \citep{Bundy2015} is one of the three projects of the fourth generation of the Sloan Digital Sky Survey \citep[SDSS,][]{York2000}. It has recently completed its goal to observe a representative sample of over 10.000 galaxies in the nearby universe ($\langle \mathrm{z} \rangle \sim 0.03$) with IFUs, obtaining a large sample of spatially resolved spectroscopic data. It uses the BOSS spectrographs \citep{Smee2013} on the 2.5 m Sloan Telescope at Apache Point Observatory \citep{Gunn2006}. The adopted IFUs consist of a set of bundles of optical fibers connected to the spectrographs, with a different number of fibers that varies from 19 fibers (covering 12" in diameter) to 127 fibers (covering 32" in diameter). These are used for different targets depending on their extension \citep{Drory2015}. The instrument also includes 12 sets of 7 fiber mini-bundles for flux calibration and 92 individual fibers used for sky subtraction \citep{Yan2016}.

The spectral data cover a wavelength range from 3600 \AA{} to 10.300 \AA{} at a resolution of R $\sim$ 2000.
The raw data was processed using the MaNGA data reduction pipeline \citep[DRP,][]{Law2016} which produces the data cubes we use in this work.
The reduced data cubes have a point spread function (PSF) of about 2.5" FWHM and spaxel size of 0.5".

The MaNGA sample is comprised of a Primary sample, which selects galaxies that can be covered by the instrument out to 1.5 effective radii (Re) and consists of $\sim$2/3 of the objects, and a Secondary data sample, which comprises galaxies covered out to 2.5 Re and contains $\sim$1/3 of the objects. Selecting the galaxies via projected size inevitably introduces a bias. More massive, physically larger galaxies have higher redshifts, with the opposite effect occurring for physically small galaxies which are located at a much closer distance.
A color-enhanced sample was added to the survey to bolster the number of galaxies observed which correspond to the more sparsely populated areas of the star forming main sequence (SFMS) diagram, such as the green valley (GV) and low mass red galaxies.

For this work, we have refined the mother sample of a little over 10,000 galaxies to suit our purpose better. First we remove galaxies that are edge-on or excessively inclined, as they would be less reliable in terms of the determination of their parameters \citep[][]{Ibarra-Medel2019}. We chose to include only those galaxies that are below 70º of inclination.
We also removed galaxies whose emission lines indicated the presence of an AGN, using the criteria from \citet{Lacerda2020}. Our refined sample contains 9087 galaxies.

\section{Analysis}
\label{sec:analysis}
The aperture of a single MaNGA fiber at the average redshift of the sample encompasses the light of thousands of stars added up. Therefore we can assume that the number of stars that contribute to each spectrum is a good sampling of the IMF. In this way, the observed spectrum is the sum of the emission of the stellar populations that correspond to the different individual episodes of star formation.

A composite spectrum of a specific region will vary depending on the star formation history (SFH) of the population, the metal composition of the ISM in which stars of different ages formed, and the present day dust extinction. The process of fitting a library of spectra to the observations to recover the composition of the underlying stellar population is called the spectral synthesis or fossil record technique, this process allows us to probe into a galaxy's past from the features left behind in the present stellar population.

\subsection{PyFIT3D}
\textsc{PyFIT3D} is a new implementation of the \textsc{FIT3D} package \citep{Sanchez2006} adopted by the pipeline \textsc{Pipe3D} \citep{Sanchez2016a,Sanchez2016b}. This code allows the user to analyze IFU data of galaxies, obtaining the full information of the ionized gas emission lines as well as applying the spectral synthesis technique to the stellar spectra to obtain spatially resolved information on the galaxy's composition and history. It is a port of the previous version to \textsc{Python 3} which makes it run $\sim 5$ times faster on average while producing consistent results (Lacerda et al., in prep.).

The process of fitting the stellar templates yields the main product that the pipeline will use to derive the physical properties, i.e., the fraction of light corresponding to each template at each spatial position. This fraction of light informs us of the current contribution that each population has toward the emission. We can correct this fraction based on the predicted loss of mass due to the stars that have died since the time the population was produced, i.e., its age. This has an effect by which the older a stellar population is the larger proportion of the stars they consist of will have died compared to the more recent ones. This is compounded with the fact that the changes to a composite spectrum due to age get less prominent for older populations. Because of this, stellar template libraries tend not to have a constant sampling on their ages with a finer sampling for the younger populations \citep[see ][]{Conroy2013,Walcher2011,Ibarra-Medel2016}.

Using the corrected fractions of mass we can obtain the number of stars formed at each time and thus, the SFH whose cumulative function is the stellar mass assembly history (MAH) \citep{Panter2003,CidFernandes2013,Garcia-Benito2017}. If we instead probe the metallicity, we can obtain the spatial distribution of metallicity values, usually studied as the radial metallicity gradient \citep{CidFernandes2013,Sanchez-Blazquez2014, GonzalezDelgado2015, Sanchez2020,Sanchez2021}. The values of the metallicity which we calculate are averages of the populations and as such they can be weighted by stellar mass or by luminosity. In this work we weigh the populations by their luminosity.
Exploring the temporal and metallicity information for stars of different ages the stellar metallicity is derived at different cosmic times, allowing us to  obtain the ChEH, which is the main property we are interested in for this article.

\subsubsection{Stellar population libraries}
The stellar population templates used for this article were constructed using the MaStar stellar library \citep[SDSS DR15,][]{Yan2019}.
The implementation of MaStar is ideal since the spectra of the stars in this library were observed with the same instrument used to obtain the MaNGA data \citep{Drory2015}.
The template of stellar populations was created using the \textsc{galaxev}\footnote{http://www.ascl.net/1104.005} \citep{Bruzual2003,Yan2010} stellar population synthesis code, assuming a Salpeter IMF \citep{Salpeter1955} and the \textsc{parsec} set of isochrones \citep{Bressan2012}. The code generates a grid of 3360 SSP templates that comprises 210 ages and 16 metallicities.
This is by far too large to be used as the base of the fitting procedure for the reasons discussed in Sanchez et al. 2016 and Lacerda et al. (submitted). From this library we select a subset of SSPs with a reasonable sampling of the age-metallicity space, as described in Sanchez et al. (in prep.)
This final library is named sLOG. It covers a total of 7 metallicities and 39 ages, consisting of a total of 273 templates. The distribution of the templates in the parameter spaces is also shown in Appendix A, in which a comparison between different samplings of the stellar ages is also presented.

As already discussed in \cite{Sanchez2018}, this library is different than the one adopted in previous explorations of the MaNGA dataset using Pipe3D \citep[e.g.][]{Ibarra-Medel2016,Sanchez2017,Sanchez2019a, Cano-Diaz2019}, and a previous exploration of the ChEH using the CALIFA dataset \citep[][, hereafter CF21]{Camps-Farina2021}. In those cases we adopted the GSD library \citep[GSD156][]{CidFernandes2013}. For completeness we have included the main figures and results discussed along this article using this former library in Appendix B. This will allow the user for a direct comparison with the results presented in CF21.

\subsection{Averaging the properties}
\label{sec:averaging}
The output of \textsc{pyPipe3D} allows us to obtain the individual ChEHs for each galaxy. Following CF21, (i) we can compare the chemical evolution of the galaxies depending on their mass, morphology, and star-forming status. In addition, (ii) we obtain the MZR at different cosmological times using the information contained in a set of ChEHs averaged within mass bins. An individual ChEH contains metallicity-time data points, by separating them into mass bins and averaging them, we have data points in the metallicity-time-mass parameter space. The MZR is immediately obtained by selecting a slice in the time dimension, yielding mass-metallicity data points at a particular time.

An average of the ChEH of a set of galaxies has to be done in a particular manner, taking into account some caveats and details (as already highlighted in CF21). One of the main concerns is that galaxies in the sample lie at different redshifts (z $\sim $ $0.01-0.15$ for the MaNGA sample), which means that the light-travel time (LTT) differs between them. When we fit the spectra of stellar population templates to the light we receive from the galaxy LTT is not taken into account (beside the redshift effect). Correcting this is trivial, we only need to add the LTT (given by the redshift via the adopted cosmology) to the ages. This, however, introduces two problems: first, the coverage in LBT is not uniform: Only nearby galaxies cover the more recent LBTs. Second, the sampling in time differs slightly between galaxies. The first caveat can be mitigated by limiting the LBT range to avoid the latest times, which are only represented by a handful of galaxies (not being representative of the full population). The second caveat can be solved by interpolating the individual ChEHs.

The averaging process itself is conducted in two steps. In the first step, we scale the ChEHs in each group (mass bins, mass and morphology bins or mass and star-formation rate bins for this work), to a characteristic metallicity of the group (the average of the currently observed metallicities). In the second step, we average the scaled ChEHs, which yields the representative ChEH for the group. The reason we do this is the non-homogeneous coverage of the LBT, which means that not all galaxies contribute to all the times in the LBT. Thus, if we performed an average at each time value using only the galaxies that contribute there, we could introduce spurious behaviors into the ChEH. As an example, consider a very low metallicity galaxy which only contributed to the earlier half of the LBT range. A simple averaging would result in the anomalous galaxy lowering the average value of the metallicity only at these times, artificially inducing a "growth" in metallicity that is not representative of the group. By scaling the ChEHs first the anomalous galaxy's effect on the average ChEH is not restricted to a specific part of the LBT range. This is a better approximation of the average that we would obtain if we had measurements for all galaxies at all times.

Another advantage of performing the averaging in this manner is that we can separate the sources of the variance within the group. The variance in the current values of the individual ChEH represents the offset between them, while the variance after shifting them is mainly due to the different rates at which they become enriched. This allows us to study whether a group of galaxies has similar growth patterns but very different values of their current metallicity or vice-versa whether their average metallicity values are similar but have been enriched at a very different rate.

It is worth noting that there is a difference in how we perform the average compared to CF21. In that article to perform the scaling rather than taking the currently observed value for each ChEH at their full LBT range we took the average value over the LBT range in which all galaxies had metallicity values. For the MaNGA sample, given the higher number of galaxies and a wider range of redshifts, this range is short and biased toward earlier LBT where most galaxies were still in the initial enrichment process. The reason for this difference between the samples is the much narrower range in redshift that the CALIFA sample covers (z $\sim $ $0.005-0.03$) compared to MaNGA (z $\sim $ $0.01-0.15$).

\subsection{Morphology classification}
\label{sec:morph}
In order to study how morphology affects the chemical enrichment we need a catalog for the morphology of the galaxies in our sample. Current publicly available catalogs only contain those galaxies included in the releases up to DR15 representing about 4700, less than half the galaxies available in the full sample.
For the purposes of this article, we do not need a highly accurate determination of the morphology for each individual galaxy, it is enough if we have a consistent classification in terms of statistical properties.

For this purpose we use the morphological classification provided by Sanchez et al. (submitted). This classification is based on a automatic machine learning algorithm that uses as a training and testing sample the morphology determinations provided for a sample of $\sim$6000 galaxies included in the SDSS VAC catalog by \cite{Hernandez-Toledo2010} \footnote{\url{https://data.sdss.org/sas/dr16/manga/morphology/manga_visual_morpho/1.0.1/}}.
\section{Results} \label{sec:results}
The analysis described before provides two main products: (i) the ChEH which shows how the metallicity has changed for a particular group of galaxies through cosmic times; and (ii) the MZR, which shows the distribution of metallicities as a function of the mass of a population of galaxies at a certain cosmological time.
In addition we obtain the variance in both the currently observed value of the metallicity and the variance in the shape of the ChEH.

\subsection{Chemical enrichment history} \label{sec:cheh}
 \begin{figure}
 	\includegraphics[width=\linewidth]{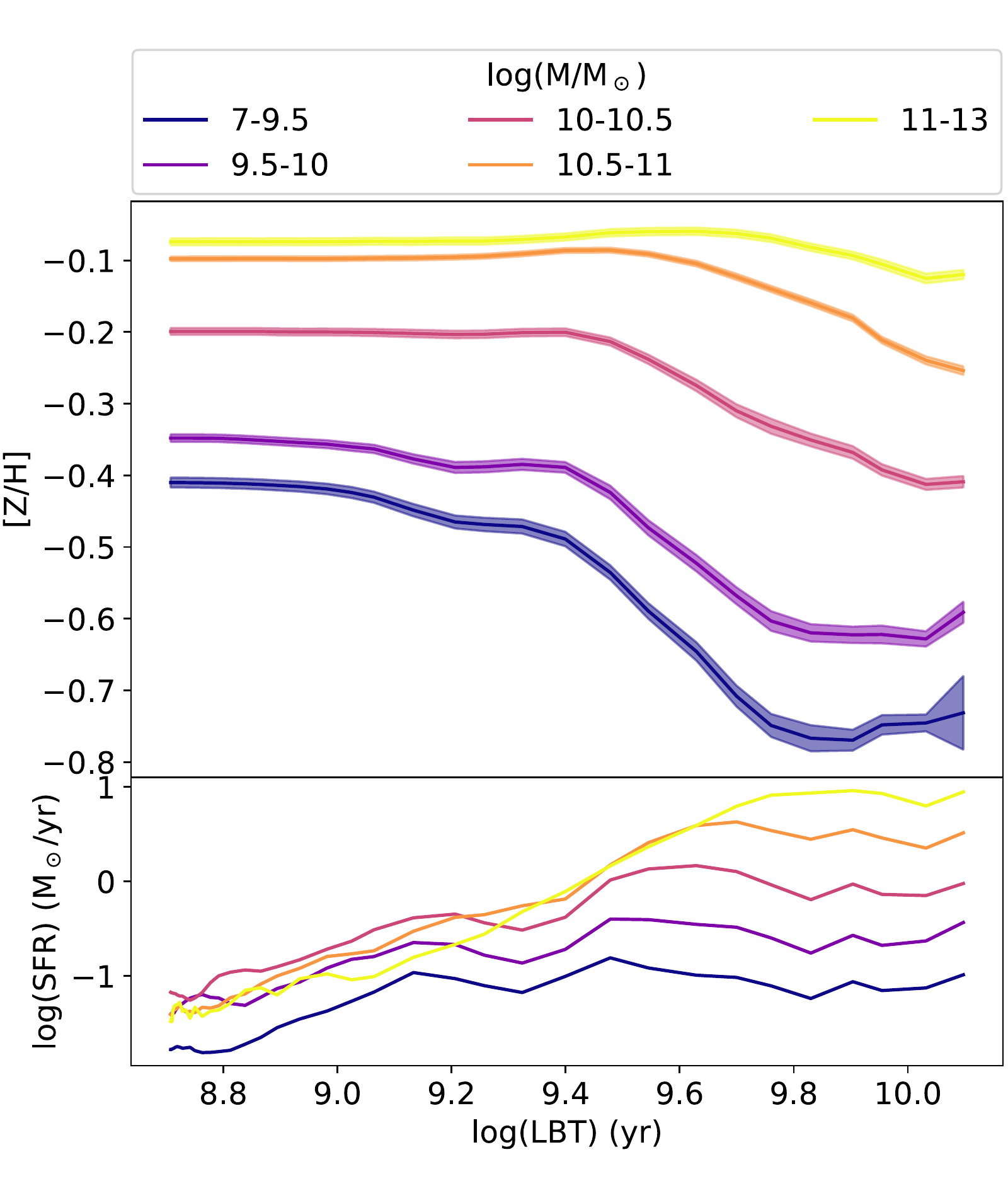}
     \caption{Chemical enrichment histories (top panel) and star-formation histories (bottom  panel) of all the galaxies in our sample binned in five stellar masses (colors). For each solid line, that represents the average distribution, the shaded areas correspond to the bootstrapped error of the mean (i.e., the range of metallicities covered within each bin has been explicitly removed).}
     \label{fig:zh_all}
 \end{figure}
In Fig. \ref{fig:zh_all} we show the average ChEH for the full sample as well as the SFH for different mass bins. We find a clear segregation of the average ChEH in the metallicity axis, with no mass bin crossing the ChEH of any other along all cosmic times. It is clearly seen that the more massive galaxies are also the ones that have the higher metallicity content at any cosmological time on average. However, this is not necessarily the case if we consider individual galaxies. It is possible for a galaxy from a lower mass bin to have higher metallicity than one from a higher mass bin.

There is also a change in the shape of the average ChEHs that varies with the mass of the galaxies. Less massive galaxies show a clear ongoing increase in metallicity, whereas the more massive galaxies have a shallower shape, almost flat for the most massive ones. This is readily apparent from how the gap between them gets narrower towards recent cosmic times.

The SFH also shows a correlation with stellar mass. The more massive galaxies have a higher SFR at earlier times. They present a steeper negative slope, and the less massive galaxies show a lower SFR along all cosmological times and a flatter distribution. This is in line with \citet{Panter2007,Perez-Gonzalez2008,Husemann2010,Perez2013, Sanchez2018, Sanchez2020} and shows how the more massive galaxies had a very high SFR at early times in contrast with low mass ones. The latter continue to steadily form stars on average.

\subsection{Evolution of the MZR}
\label{sec:mzr}

 \begin{figure}
 	\includegraphics[width=\linewidth]{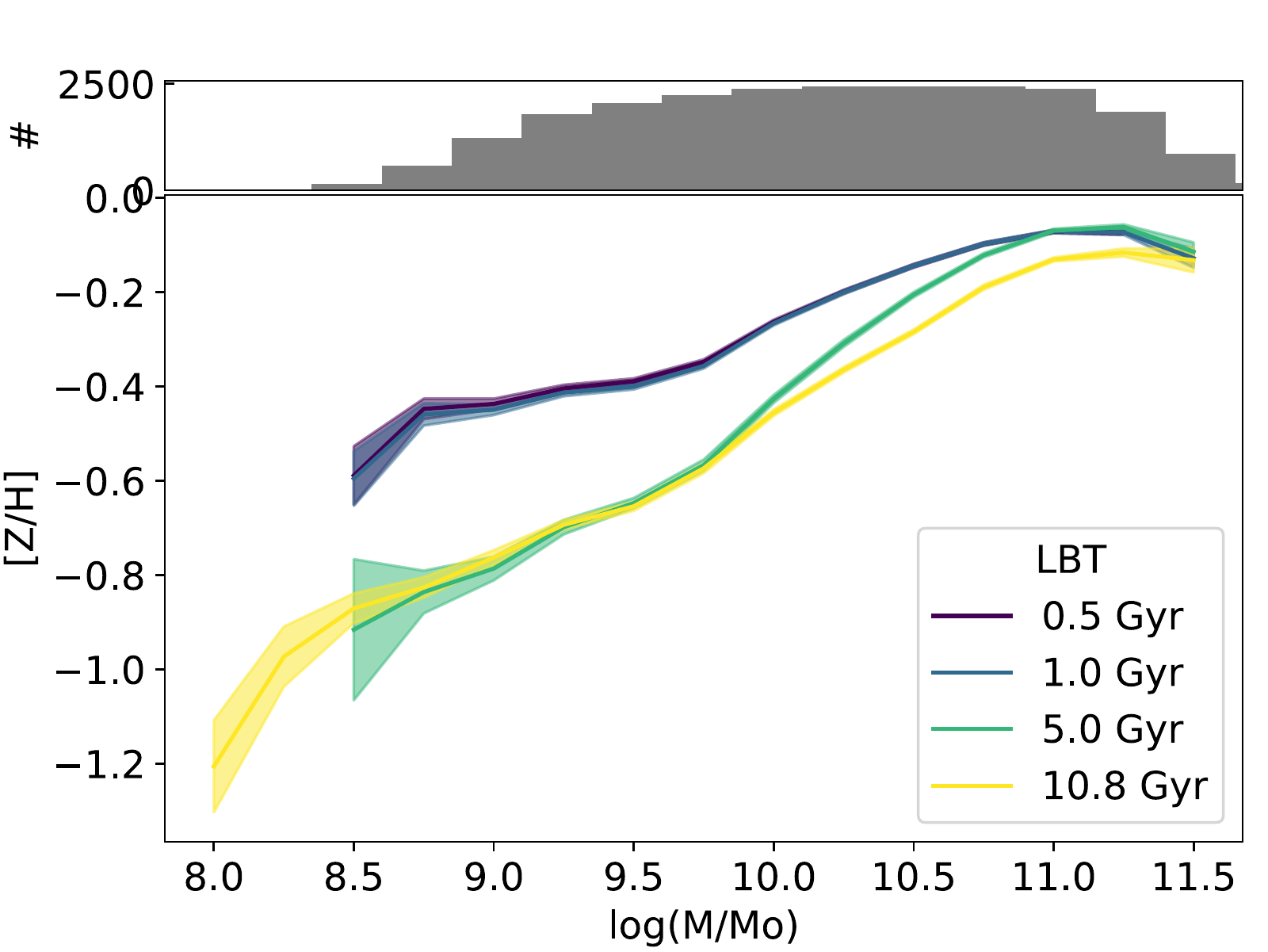}
     \caption{Evolution of the MZR along the cosmological time (bottom panel) and stellar mass distribution of the galaxies in our sample as currently observed (top panel). Shaded areas in the bottom panel represent the bootstrapped error for each ChEH derived by the averaging algorithm.}
     \label{fig:mzr_all}
 \end{figure}
 
In Fig. \ref{fig:mzr_all} we show the MZR at different cosmic times.
The main features observed are consistent with those observed in Fig. \ref{fig:zh_all}.
The slope of the MZR is always positive towards higher mass showing that, at all times, the more massive galaxies have higher metallicities. The value of the slope itself, however, changes along cosmological time, becoming shallower towards recent times. There is a clear delay in the enrichment due to the mass, observed especially around the 5 Gyr MZR. At this time the most massive galaxies have already reached their metallicity asymptotic value. On the other hand, low mass galaxies have not significantly increased their metallicities at this time compared to the higher LBTs sampled (i.e., between 5 and 11 Gyrs).
This is a consequence of high mass galaxies becoming enriched earlier than low mass ones. This result is in line with what we found in the previous section, with the higher mass galaxies having shallower ChEHs. High mass galaxies became enriched very early on, while lower mass galaxies have steadily been catching up to them, thus producing the observed change in slope.

The MZR shows a flattening towards high masses which would imply a kind of saturation regarding the enrichment of the medium. This flattening has been observed before in the nebular MZR \citep[][]{Tremonti2004,Kewley2008,Rosales-Ortega2012,Sanchez2014,Barrera-Ballesteros2017,Sanchez2019a, Blanc2019}. Among the proposed explanations for this observational result the ones with the strongest support at those related to the effect of metal-rich outflows and metal-poor inflows in shaping this distribution \citep[e.g.,][]{Gallazzi2005,GonzalezDelgado2014,SanchezAlmeida2014,Barrera-Ballesteros2017,Sanchez2018,Lacerda2020}. \citet{Zahid2014} proposes that the flattening is the result of the saturation of the metal content in the gas. This occurs once the metallicity reaches a value such that the metal mass locked-up in newly formed low mass stars is similar to the metal mass produced in newly formed massive stars.

The low mass regime, below $\sim10^{9.7}$ M$_\odot$ also appears to flatten, especially at recent times, though to a lesser extent than for the high mass range.
This feature has been observed previously in \cite{Gallazzi2005} for stellar metallicity and in \cite{Kashino2016,Blanc2019} for nebular metallicity. It is also consistent with some high resolution galaxy simulations \citep[][]{Schaye2015, Ma2016,Christensen2016, Christensen2018} which show a flatter MZR for the low mass regime under $10^{9.5}$ M$_\sun$.
There are, however, other studies in nebular metallicity in which the flattening is not observed such as \citet{Lee2006,Berg2012,Zahid2012}. \citet{Blanc2019} discusses these discrepancies and how consistent this feature is, which they attribute to a characteristic mass ($10^{9.5}$ M$_\sun$) above which the efficiency of metal-removal processes drops.

One thing to note is that the mass at which the MZR steepens does not appear to change with LBT, which is especially apparent comparing the 5 and 10.8 Gyr MZR. The two are practically the same up to $\sim 10^{9.8}$ M$_\sun$ at which point the 5 Gyr MZR steepens. The fact that the turn-up mass remains constant in time supports the hypothesis that it is the characteristic stellar mass at which the gas removal processes become less efficient.

Another explanation for the flattening that must be taken into account for our study compared to others such as \citet{Tremonti2004} is that we are using stellar metallicity instead of nebular one. We obtain our values from the fitting of stellar population templates and therefore we are limited to the metallicity values that we can obtain by the parameter space covered by the templates. This can potentially produce a flattening of the MZR as the measured values of the metallicity approach the edge values of the stellar template library. The figures in Appendix \ref{sec:appendix-libraries} and \ref{sec:CALIFA} are helpful to assess whether this is the case for our results, as they show the MZR obtained by using a different stellar template library which has a narrower range in metallicity compared to MaStar. In general, the narrower metallicity range creates quantitative differences but not many qualitative ones. There does not appear to be a saturation effect such that bins with low metallicity cannot be distinguished, though the differences do become less clear.

\subsection{Effect of morphology}
\label{sec:morph_results}
 \begin{figure*}
 	\includegraphics[width=\linewidth]{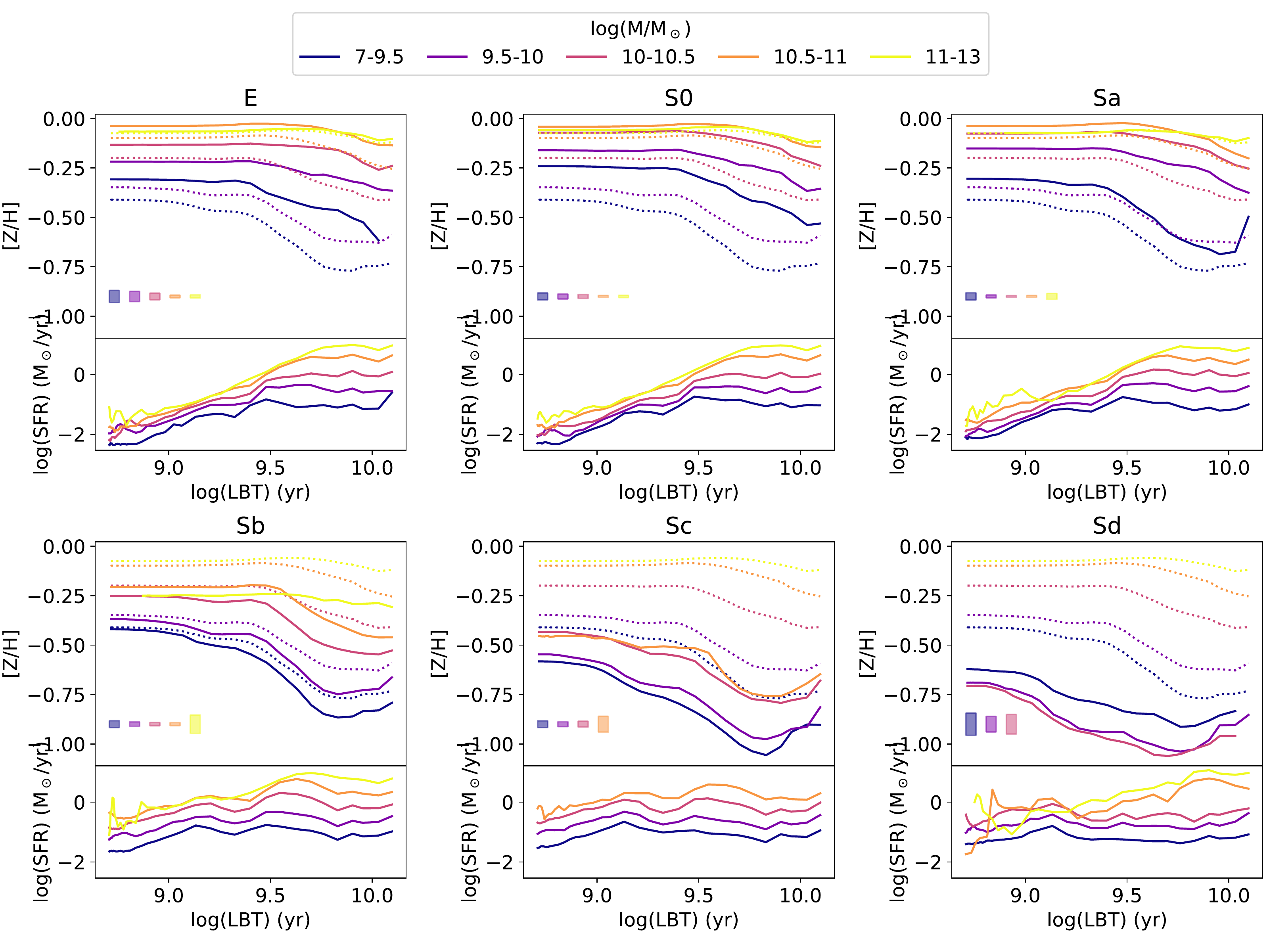}
     \caption{Chemical enrichment histories (top of each panel) and star-formation histories (bottom of each panel) of galaxies in our sample segregated by morphology. The dotted lines correspond to the ChEHs from Fig. \ref{fig:zh_all} which represent the full sample, shown here for the sake of comparison. The shaded areas show the average bootstrapped error for each ChEH, in the same colors.}
     \label{fig:zh_morph}
 \end{figure*}

  \begin{figure*}
 	\includegraphics[width=\linewidth]{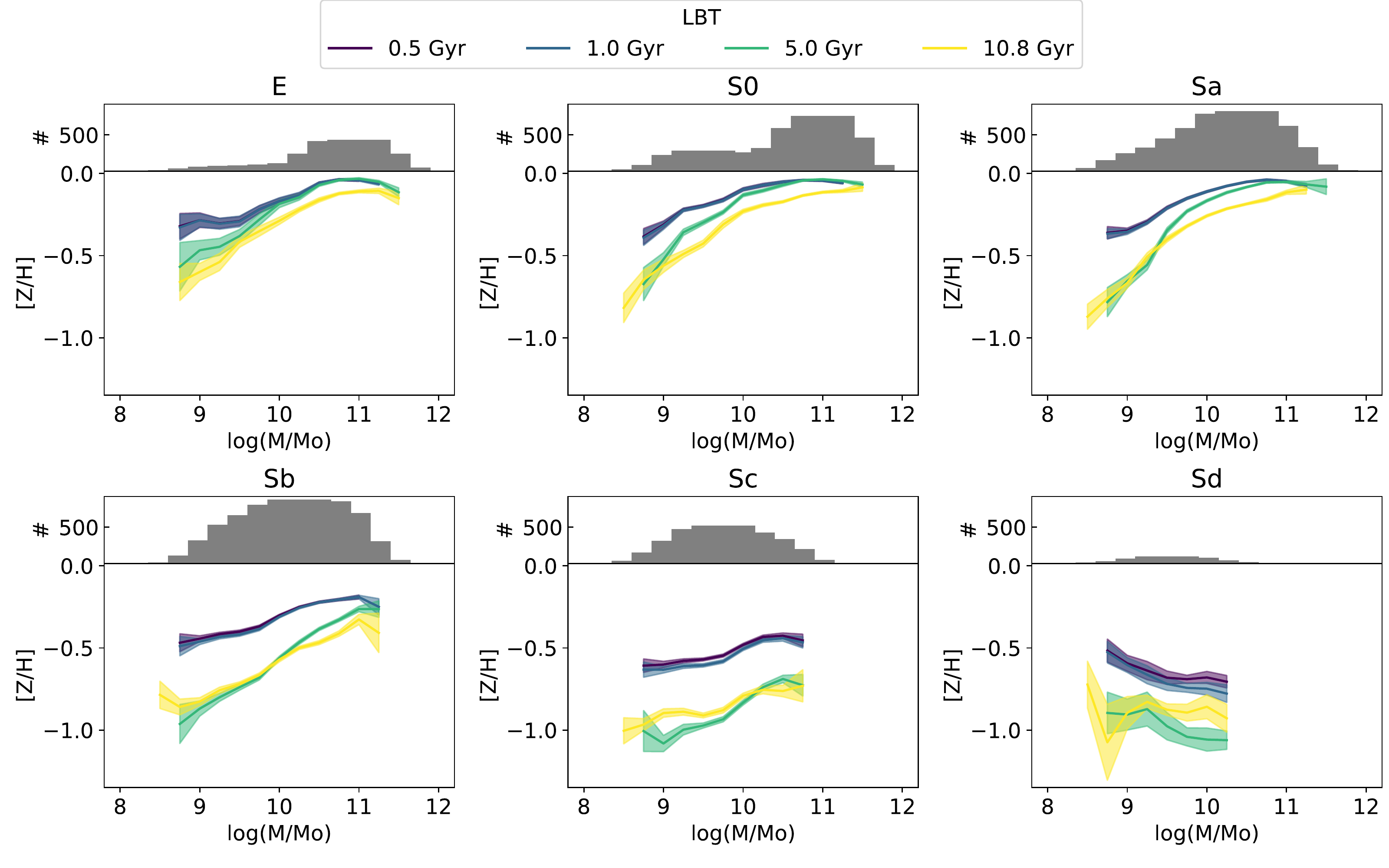}
     \caption{Evolution of the MZR along the cosmological time for all galaxies in our sample separated by morphology. On the top panel of each plot we show the distribution of galaxies along the mass range as currently observed. Shaded areas in the bottom panel represent the bootstrapped error.}
     \label{fig:mzr_morph}
 \end{figure*}
 
We explore how the morphology affects the chemical evolution in galaxies using the classification described in Sec. \ref{sec:morph} to probe how the morphology of a galaxy affects its chemical evolution. By separating galaxies in both mass and morphological bins we avoid the effect introduced by the mass-morphology correlation.

In Figs. \ref{fig:zh_morph}, \ref{fig:mzr_morph} we show the equivalent of Figs. \ref{fig:zh_all}, \ref{fig:mzr_all} but separated into morphology bins. In Fig. \ref{fig:zh_morph} we can appreciate that earlier type galaxies are generally more metallic than their late-type counterparts at a fixed mass. The same result is observed when exploring the shape of the ChEH. E galaxies have practically flat ChEHs for most mass bins, whereas spirals still present growth even at high masses (except for the M$^{11-13}$ M$_\odot$ bin), on average. More massive galaxies show this change better than lower mass ones, which are more similar in shape across morphology bins in general, especially from Sb to Sd. However, the low mass bins of E-S0-Sa galaxies have higher metallicities than later types at the same mass bin.

This effect is mirrored in the SFH distributions, with the earlier type galaxies having steeper slopes as they stopped forming stars early on. In contrast, later type galaxies have flatter profiles showing sustained star formation.

The MZR plots show results consistent with the ChEH ones, with earlier types having higher values of metallicity in general. However, the most interesting part of this figure lies in the shape of the MZR, which gets progressively flatter towards later types to the extent that Sd galaxies appear to have an inverted MZR. This can also be observed in the ChEHs, as the lowest mass bin shows a higher metallicity than the higher mass ones.
The implication of this flattening of the slope depending on morphological type is that for late-type galaxies the stellar mass is progressively less important in determining the current metallicity than the morphology. For Sd (and Sc to a lesser extent) galaxies mass appears to have little impact on the metallicity enrichment, on average.

The evolution of the MZR also changes between morphological types. The earlier types (E, S0 and Sa) become shallower as time passes and also show a delay in enrichment for lower mass galaxies. The Sd galaxies, on the other hand, show neither a change in slope or a delay in enrichment at all. The change in slope is caused by the delay in enrichment, and the fact that it is not observed in Sb and Sc galaxies indicates that the "transitional" LBT where high mass galaxies are already enriched but low mass ones are not ($\sim 5$ Gyr for the full sample) is shifted to more recent times for later type galaxies. This shows how the delay in enrichment between galaxies is two-fold: there is a delay due to mass and also one due to morphology. Thus, for two galaxies with equal stellar mass the earlier type one will become enriched at a higher LBT. On the other hand, for two galaxies of the same morphological type the more massive one will become enriched at a higher LBT.
The Sd galaxies, however, show no change in slope and therefore have no delay in enrichment due to mass.

Considering that earlier types dominate the high mass range and later types the low mass one, this implies that the global change in shape seen in Fig. \ref{fig:mzr_morph} is not simply a result of mass regulating how fast galaxies evolve. Morphology appears to be at least as important.

\subsection{Effect of star-forming status}
\label{sec:sfs} 
 Next, we explore how the star-forming status (SFS) relates to the evolution of the chemical enrichment.
We define the SFS of our galaxies based on the equivalent width (EW) of the average H$\alpha$ emission line, defined at the effective radius.
This quantity has been proven to be a good discriminator of the SFS \citep{Stasinska2008,Sanchez2014,Cano-Diaz2016,Espinosa-Ponce2020,Lacerda2020,Sanchez2020,Sanchez2021}, where an EW$_{\mathrm{H}\alpha}$ value of 6\AA{} is shown to be a good value to separate star-forming galaxies from retired ones. In this article, we want to separate into star-forming (SFG), green valley (GVG), and retired (RG) galaxies, so we take discriminating values as done in \citep[][]{Lacerda2020} of:
\begin{itemize}
\item SFG: EW$_{\mathrm{H}\alpha}$ (R$_\mathrm{e}$) $>$ 10\AA{}
\item GVG: 3\AA{} $<$ EW$_{\mathrm{H}\alpha}$ (R$_\mathrm{e}$) $<$ 10\AA{}
\item RG:     EW$_{\mathrm{H}\alpha}$ (R$_\mathrm{e}$) $<$ 3\AA{}
\end{itemize}

 \begin{figure*}
 	\includegraphics[width=\linewidth]{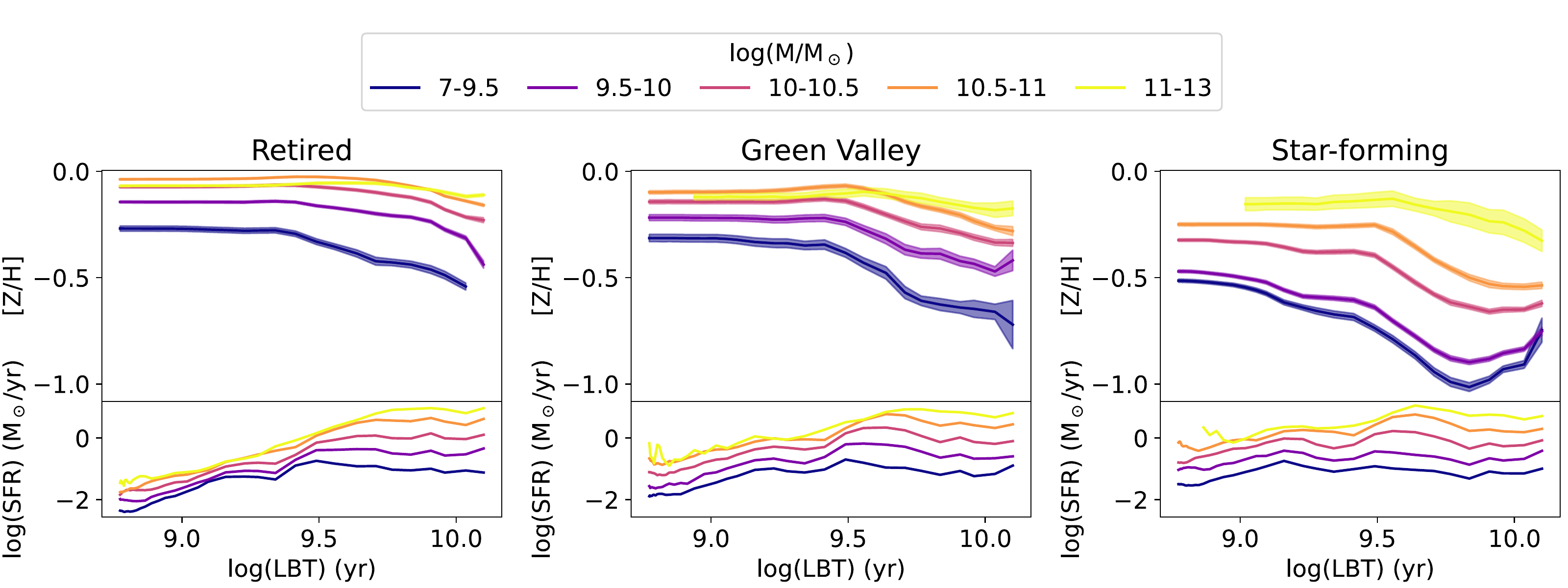}
     \caption{Chemical enrichment histories (top panel) and star-formation histories (bottom  panel) of galaxies in our sample segregated by their current star-forming status.  Shaded areas in the top panel represent the bootstrapped error.}
     \label{fig:zh_sf}
 \end{figure*}
 
  \begin{figure*}
 	\includegraphics[width=\linewidth]{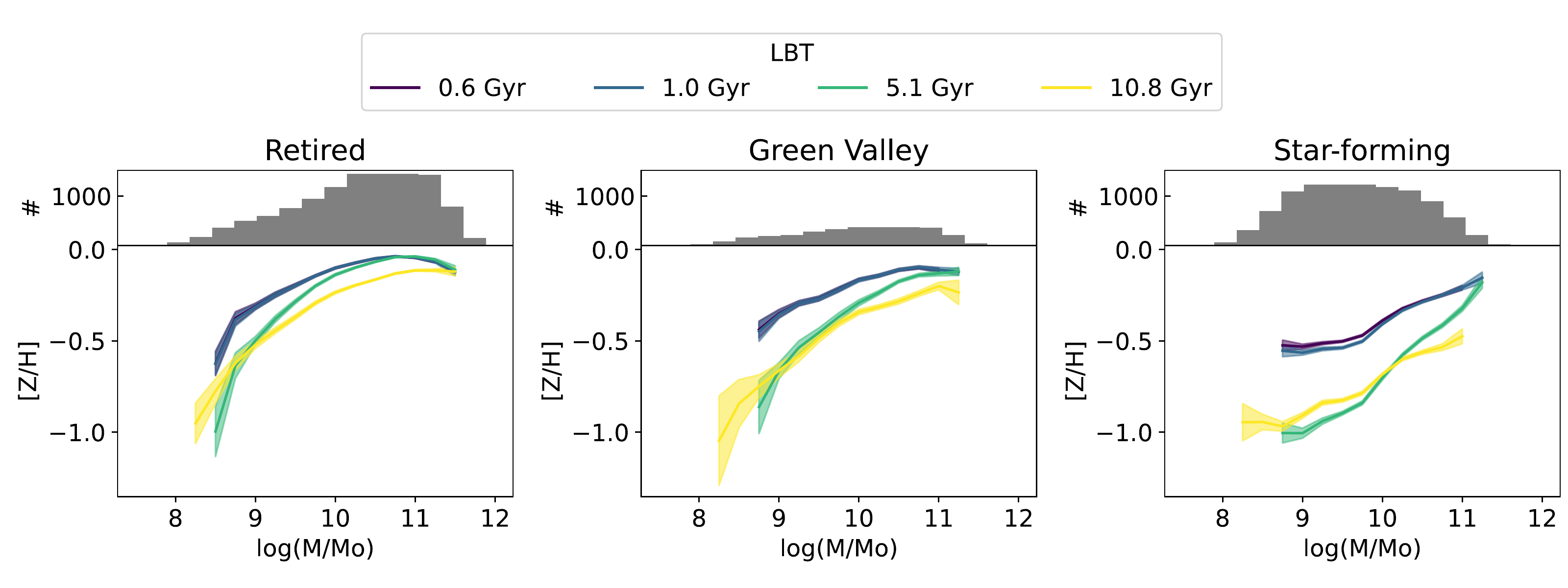}
     \caption{Evolution of the MZR along the cosmological time (bottom panel) and stellar mass distribution as currently observed (top panel) for all galaxies in our sample segregated by their current star-forming status. Shaded areas in the bottom panel represent the bootstrapped error.}
     \label{fig:mzr_sf}
 \end{figure*}

In Figures \ref{fig:zh_sf}, \ref{fig:mzr_sf} we show how the SFS affects the results. It is important to take into account that this separation is not independent of the morphology. As seen in the previous section, late-type galaxies are more likely to have maintained a high SFR at current times, whereas early type galaxies are likely to be retired.
To claim that a particular feature is caused by the morphology or the SFS of galaxies in a group, the feature needs to be unique within its equivalent morphology and SFS bins. In other words, only a feature that (for example) appears in the SFG but not in late-type galaxies could be clearly claimed to be a result of SFS.

In this manner, we can see many expected results. SFG are generally less metallic and have steeper shapes in their ChEHs than GVG and RG, similarly to their dominant morphological components.
An interesting feature that SFG show is a steeper enrichment between 10$^{9.3}$ to 10$^{9.7}$ yr, a feature that differs from RG and GVG. This slope change is more important for higher mass galaxies, with the two lowest mass bins barely showing a change.

The MZR also shows the expected changes between SFS bins inherited from their morphology, with retired galaxies being more metallic than GVG and especially SFG. The RG show a much more prominent flattening at high masses compared to GVG, with SFG showing no high mass flattening at all. The opposite happens for the low-mass flattening which is absent in RG and GVG but clear in SFG. The evolution also shows differences: The delay in enrichment, which produces the change in slope, is more prominent in RG than in GVG and is absent in the SFG for the shown LBT values. As in Section \ref{sec:morph_results}, this implies that the transition time is closer to us for GVG and especially for SFG compared to RG.

Many of these features correlate with the corresponding ones in morphology, but some features cannot be explained only through the SFS-morphology correlation. The low mass flattening of the MZR is more prominent in the SFG than for any morphological bin. It could be argued that it is an effect of Sd galaxies dominating the lowest mass range, but if we check the top panels of Figure \ref{fig:mzr_morph} we can see that Sd galaxies are too few to dominate over Sb even at the low mass range. The evolution of the MZR for SFG is more similar to that of Sb galaxies. This is consistent with their numbers and likely-hood of being in the SFG bin, but with a steeper slope and a more prominent low-mass flattening. The steeper slope can be explained as the effect of adding Sa galaxies, but the flattening appears to correlate more with the SFS, rather than being induced by the morphology.

The opposite happens for the high mass flattening. This feature can be observed clearly for Sa galaxies and to a lesser extent also for Sb and Sc ones, though for the latter the MZR is fairly flat overall. However, it is absent entirely in the SFG bin. Sa and Sb galaxies should be the dominant morphological types (especially at high stellar masses) in the SFG, so this clearly implies that the high mass flattening of the MZR is strongly related to the SFS of galaxies. This link does not need to be direct, following \cite{Zahid2014} (see \ref{sec:mzr}) the flattening would be the consequence of RG and GVG having already reached the equilibrium metallicity while most SFG have not done so.

\subsection{Effect of radial distance}
\label{sec:radius}
Another parameter we can explore is the radial distance at which we measure the metallicity. So far, we have shown results measured at the Re, which is a good proxy for the global metallicity of a galaxy \citep[e.g.,][]{GonzalezDelgado2014,Sanchez2020} and is, therefore, a good scaling quantity to allow us to average galaxies of different sizes.

Using the slope of the metallicity gradient, which is a parameter obtained from the \textsc{pyPipe3D} pipeline, we can infer both the metallicity at the center and the outskirts, which we have taken as 2 Re. This inference is only strictly valid if the gradient is linear, as the slope fitting algorithm assumes. However, we consider it a good proxy for how the metallicity changes at different radial distances.

 \begin{figure*}
 	\includegraphics[width=\linewidth]{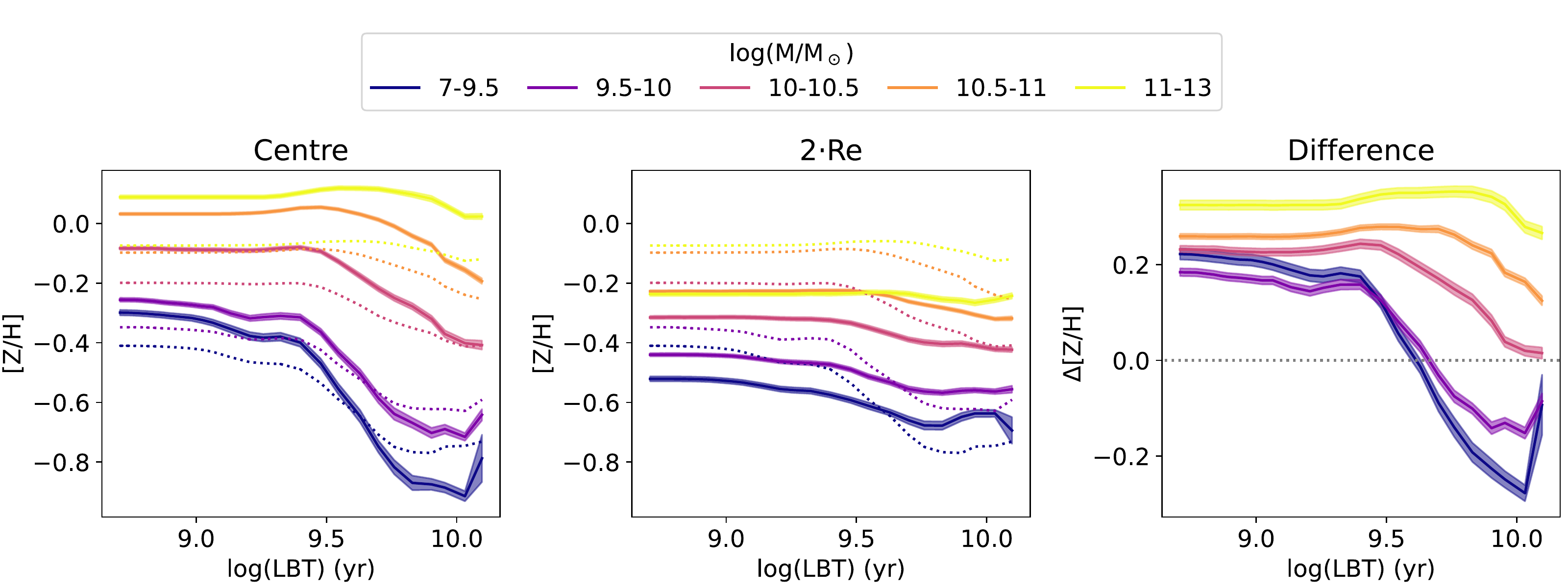}

     \caption{Chemical enrichment histories of galaxies in our sample measured at the center (left panel) and at twice the effective radius (center panel) as well as the difference between them (right panel). Shaded areas represent the bootstrapped error. The dotted lines in the left and center panels correspond to the ChEHs at the effective radius.}
     \label{fig:zh_pos}
 \end{figure*}

 \begin{figure*}
 	\includegraphics[width=\linewidth]{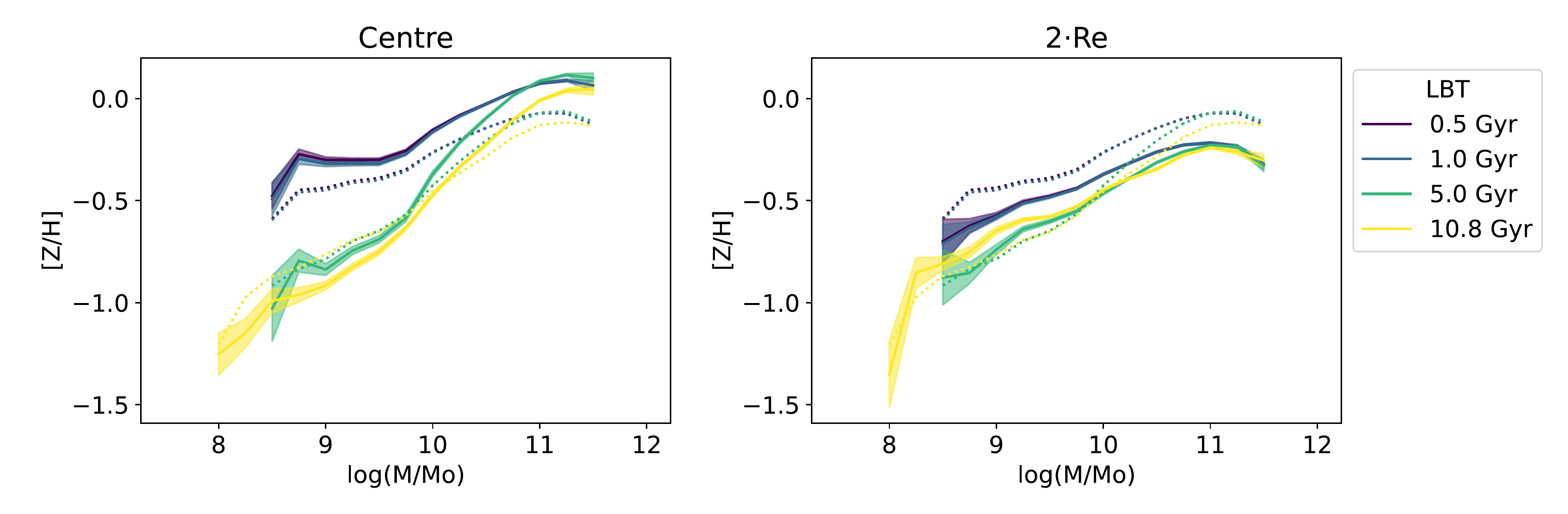}
     \caption{Evolution of the MZR along the cosmological time for all galaxies in our sample measured at the center (left panel) and at twice the effective radius (right panel). Shaded areas represent the bootstrapped error. The dotted lines in the left and center panels correspond to the MZR at the effective radius.}
     \label{fig:mzr_pos}
 \end{figure*}

In Figs. \ref{fig:zh_pos},\ref{fig:mzr_pos} we show the ChEHs and the MZR which result from measuring the metallicity at the center and at 2 Re. In general terms, the center tends to be more metallic and has a more pronounced evolution, whereas the ChEH has a similar shape for all mass bins at the outskirts. The gap between the ChEHs is also narrower. This is consistent with local downsizing, as the outskirts of galaxies are more similar in surface stellar mass density ($\Sigma_*$) compared to the center. Local downsizing is the extended version of downsizing, establishing a correlation between $\Sigma_*$ and how fast a region's stellar population evolved, such that the more dense regions evolved faster than the lower density ones. This is similar to how the more massive galaxies assembled their stellar mass faster than low mass ones.\citep{Perez2013, Ibarra-Medel2016, Garcia-Benito2017}

The MZR shows similar behaviors, with the change in shape over cosmic time being more evident at the center but practically non-existent at the outskirts. Generally, there are lower values of the metallicity and a flatter shape, which indicates that the differences in metallicity due to galaxy mass are less prominent. This shows once again that in their outskirts galaxies are more similar for different stellar masses in terms of their chemical enrichment.

A particularly interesting feature can be seen in the ChEHs for lower galaxy masses. For recent cosmic times, all mass bins show a higher metallicity in the center rather than the outskirts, a feature that is maintained at all LBT for the high mass bins. For masses below $10^{10}$ M$_\sun$, on the other hand, the opposite trend is seen at earlier times with a higher metallicity in the outskirts rather than the center. As the stellar mass decreases, the positive gradient at early cosmic times is higher, and the switch to inside-out growth is done at more recent cosmic times. This implies that the lower mass galaxies shifted from an outside-in growth in metallicity to an inside-out one approximately 4 Gyr ago (10$^{9.6}$ yr). It bears mention that the specific time for this inversion is likely to be heavily affected by the model for the stellar templates, so it should not be taken as a precise measurement \citep[][]{Ibarra-Medel2019,Sanchez2020}.
\citet{Hidalgo2013} finds that for four isolated dwarf galaxies, the stellar populations suggest an initial outside-in scenario after which the SFR is quenched towards the center as the gas at the outskirts runs out. While these galaxies lie below our mass range at M$_{*} = 10^{6-7}$ M$_\odot$, their behavior is consistent with our results.

Another interesting result can be seen in the flattening of the MZR at high and low masses. The high mass flattening appears both at the center and the outskirts, but it is more prominent in the latter. The low mass flattening, on the other hand, is only clearly observed at the center and for recent times. Below 10$^{9.5}$ M$_\odot$ and up to 1 Gyr in LBT the MZR is completely flat at the center.
This can be seen to a lesser extent in Figure \ref{fig:zh_pos} by comparing how the ChEH of the two lowest bins converge more at recent times in the center than in the outskirts.

\subsection{Variances}
\label{sec:variance}
Due to the unique way in which we have performed the averaging of the ChEHs, as described in Section \ref{sec:averaging}, we can separate between two sources of variance that contribute to the standard deviation of the averaged ChEHs. One is dependent mostly on the current value of the metallicity for each ChEH, which we name $\sigma_{Scale}$, and the other is related to how diverse are the shapes of the ChEHs, or the rate of enrichment. The second one we name $\sigma_{Shape}$.

 \begin{figure}
 	\includegraphics[width=\linewidth]{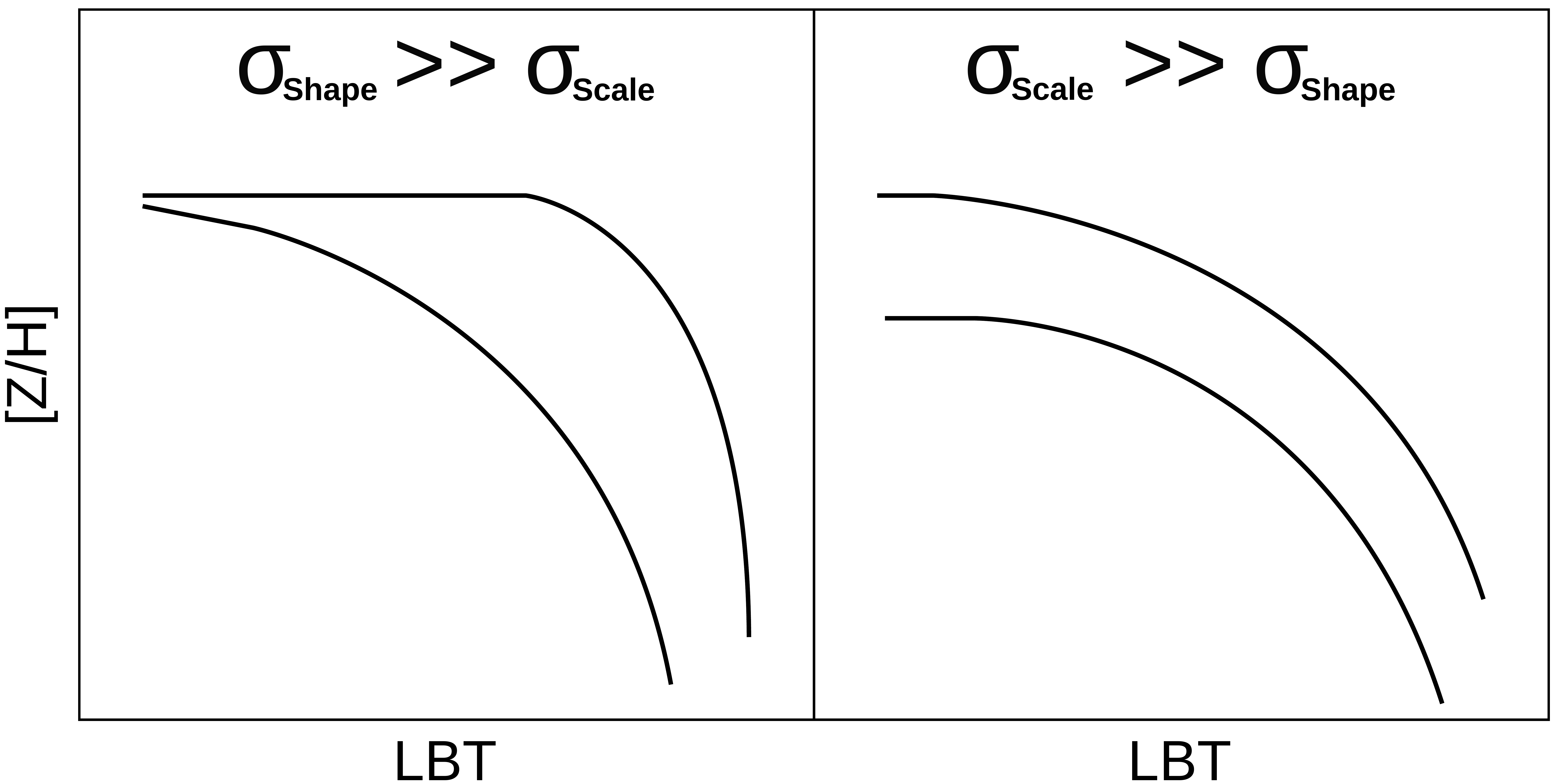}
     \caption{Diagram illustrating the difference between the two types of variance considered. In the left panel we show an example of two ChEHs with similar current metallicity but different enrichment rates, whose average would have a higher $\sigma_{Shape}$. In the right panel we show the opposite case with two ChEHs that are similar in their rate of enrichment (their "shape") but have different values of the metallicity, such that the average of the two has a higher $\sigma_{Scale}$.}
     \label{fig:sigmas}
 \end{figure}

In Fig. \ref{fig:sigmas} we show a diagram which illustrates the meaning of the two types of variance for the ChEHs.

Comparing how the values of the standard deviations vary for the different groups considered in this article gives us a direct insight into the diversity in ChEHs and how it is affected by mass, morphology, and star-forming status.

 \begin{figure*}
 	\includegraphics[width=\linewidth]{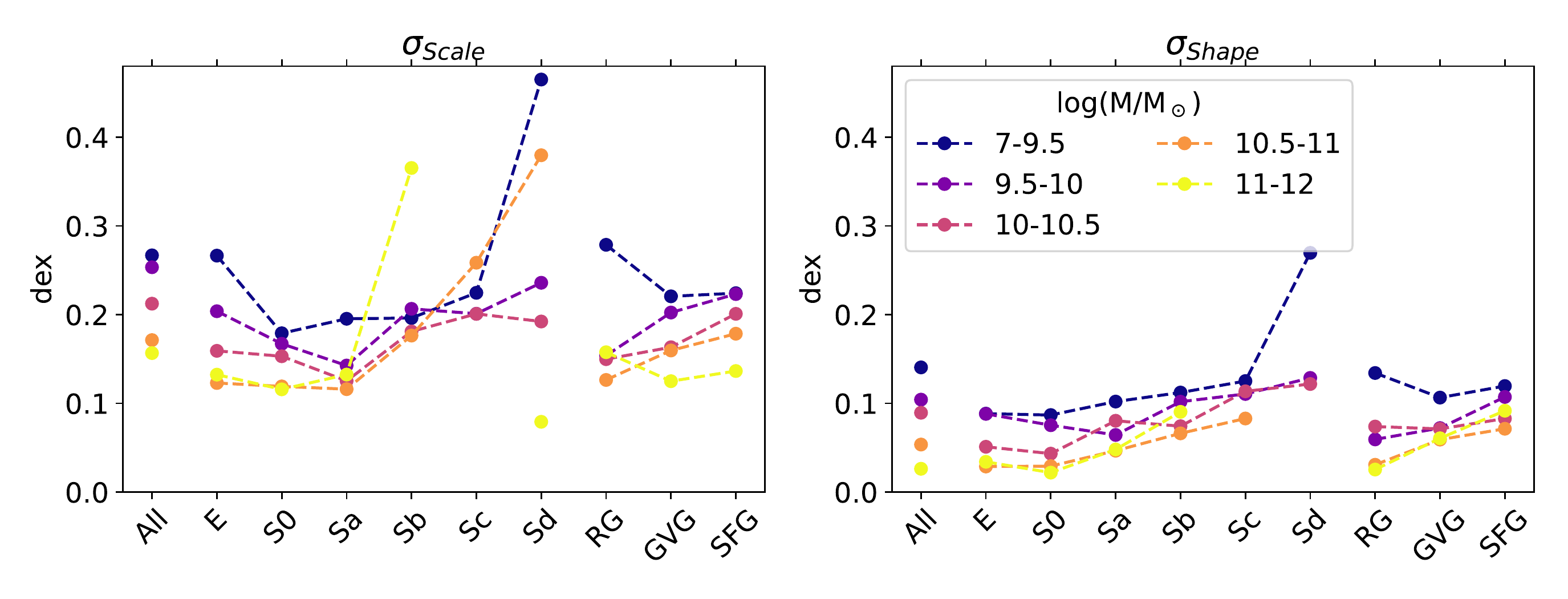}
     \caption{Comparison of the standard deviations of the ChEH for the different bins considered in this article, separated into the standard deviation due to the differences in the absolute value of the metallicity ($\sigma_{Scale}$) in the left panel and that due to a difference in the shape of the ChEH ($\sigma_{Shape}$) in the right panel.}
     \label{fig:variance}
 \end{figure*}

In Fig. \ref{fig:variance} we show the two defined components of the variance for the galaxies in our sample, divided by morphology, mass, and star-forming status. The first result that can be observed is that $\sigma_{Scale}$ is generally higher than $\sigma_{Shape}$. This implies that within any classification scheme, the differences between ChEHs are primarily due to absolute differences in the value of the metallicity rather than differences in enrichment rate.

Both $\sigma_{Scale}$ and $\sigma_{Shape}$ tend to increase from early morphological types towards late types, implying that earlier type galaxies present a more similar evolution in general. For $\sigma_{Shape}$ in particular, this is also true for more massive galaxies compared to less massive ones, an effect that is not as clear for $\sigma_{Scale}$.

When it comes to the SFS of the galaxies, the SFG tends to be more diverse than the RG. This is in line with the inter-dependency of SFS and morphology. A unique effect is observed for the dependency on mass: the correlation between $\sigma_{Shape}$ and SFS disappears for the lowest mass bin. This result can be interpreted from either the mass or the SFS point of view. In the former, we conclude that low mass galaxies have a wide distribution in terms of the shape of their ChEH, regardless of their SFS. In the latter, we conclude that the RG have a significant correlation between their mass and how diverse the shape of their ChEH is. The SFG, on the other hand, have similar enrichment rates regardless of mass compared to the RG. Naturally, both interpretations are fully compatible, as they are simply different ways to view the same result.

\section{Discussion}
\label{sec:discussion}
Among the results presented in this article there are three that we consider to be especially relevant: How morphology shapes the MZR, the correlation between star-formation and the flattening of the MZR and the inversion of the metallicity gradient for low-mass galaxies.

It is well known that stellar mass is key in determining the shape of the MZR \citep[e.g.,][]{Tremonti2004} but the role of morphology has not been consistently studied. Our results show that the global shape of the MZR is heavily influenced by morphology, with the low mass range being similar to the MZR of Sd galaxies and the high mass end being similar to E-Sa galaxies. This, of course, is not to deny the role of stellar mass: For all morphological types except perhaps Sd stellar mass is still a strong regulator of how much and how quickly galaxies become enriched. The effects vary between morphological types though. At the same stellar mass, an early type galaxy will achieve a higher metallicity than a late type galaxy, and will do so more quickly. It appears, then, that both stellar mass and morphology are essential to understand how galaxies become enriched.

The correlation we find between the SFS of galaxies and whether the MZR shows either a high or low-mass flattening suggests that the processes that sustain or quench star-formation are key in determining the chemical enrichment. The overall shape of the MZR has been generally linked to a correlation between the capacity for a galaxy to expel metals through outflows and the stellar mass as a proxy of the halo mass \citep[e.g.,][]{Tremonti2004}. In this scenario, the high-mass flattening occurs because above a certain stellar mass outflows can no longer efficiently remove gas from the galaxy.
Our results suggest that as long as a high SFR is sustained the outflows can still efficiently remove metals and it is only after the galaxy starts to become quenched that the higher potential curbs metal loss.
The alternative scenario by \citet{Zahid2014} is also supported by our results. It describes the high-mass flattening as bieng the consequence of an equilibrium being reached between the production of metals by massive stars and the locking up of metals by low mass stars. In this scenario the RG and GVG would have reached this point earlier as a result of an earlier growth and lack of other mechanisms diluting the ISM.

Alternatively, hierarchical galaxy formation models can reproduce the MZR without requiring that metal loss determines its shape, but as a result of the star formation efficiency varying with mass \citep[e.g.,][]{deRossi2007,Finlator2008}. In these scenarios our results would imply that as galaxies become quenched the relation between mass and star formation efficiency changes. This would explain both the high and low-mass flattening but this change is difficult to determine using observational data.

Regarding the inversion of the metallicity gradient for lower mass galaxies, there are high-redshift observational studies which find a substantial population of galaxies with a positive gradient \citep[][]{Cresci2010,Troncoso2014,Carton2018,Wang2019, Wang2020, Simons2020,Sharda2021} which does not appear in the local universe. One explanation for this is that the lower metallicity at the center is the result of merger-driven inflows of gas which enhance star-formation at the center and simultaneously dilute the enriched gas there \citep[see review by][]{Maiolino2019}.

Our results are compatible with the observations as well as the interpretation, though a more detailed analysis of the stellar populations is needed to confirm the latter. The SFR at the center should be significantly enhanced for the LBT at which the gradient is inverted compared to the rest of the galaxy. This is necessary in order for merger-driven inflows to produce the apparent inversion in gradient rather than it being the result of secular evolution.

This article constitutes an expansion of the work performed in the recent article CF21 for the CALIFA sample, but this time using the larger MaNGA sample and a different set of stellar templates for the fit. It is important to compare the results of these studies, as the differences in sample, instrument and method might be introducing artifacts into the results. For ease of comparison between samples we provide in Appendix \ref{sec:CALIFA} the same figures shown in this work but using the GSD stellar library. Note that the range in metallicity values shown in these figures is not the same as that of the figures for the main results. The GSD library covers a narrower range of metallicity values (see Figure \ref{fig:zh_comp}). The analysis for the MaNGA sample with GSD is mostly equivalent to that of CF21, with some differences as described in Section \ref{sec:averaging}.

The results with GSD are similar to CF21 both in terms of the absolute values of the metallicity and the shapes of the ChEH and MZR, with some notable exceptions. In CF21 Sa and S0 galaxies do not show a delay in enrichment in the MZR, unlike this work. A careful comparison of the figures shows the reason for this: the change in slope only appears if we include the mass range below 10$^{9.5}$ M$\sun$, which is not represented in the CF21 version of this figure. The reason for this is that the selection criteria of CALIFA prefers objects with a higher inclination for low mass disc galaxies, as they show a higher surface brightness. Our inclination criteria then removes most of these galaxies. E galaxies, which are populated in both samples, do show this effect.
Another difference is the slope of the MZR for Sd galaxies, which appears flat in the CALIFA sample but becomes negative for MaNGA galaxies. The progressive flattening of the MZR for later morphological types is similar between the two but appears to be more dramatic in MaNGA.

One of the standout results in CF21 was that the SFG showed a clear convergence in the ChEHs. For the MaNGA sample this effect is not as clear or distinct from the RG and GVG. It can still be observed to a degree for the three most massive bins (the two lowest ones have a steep growth for all SFS which naturally produces a convergence towards the more massive bins) in that the gap between them is narrower at more recent cosmological times than in the past, whereas for RG and GVG the gap between mass bins is constant.

The similarities of the results when using the same methodology implies that the differences between the CALIFA and MaNGA surveys (spatial and spectral sampling, sample selection, wavelength range...) are not critical for the determination of the composition of the underlying stellar populations using the spectral synthesis technique.
The methodology, on the other hand, plays a key role in determining the age and metallicity of the populations, with the choice of stellar library being paramount \citep[][]{CidFernandes2014}.
This does not mean that using different libraries will drastically affect the general behaviors, at least qualitatively. Independently of whether we use GSD or MaStar we observe that the more massive galaxies are both more metallic and become enriched faster than lower mass galaxies. The same general conclusions are reached regarding early type galaxies versus late-type ones.

The greatest differences between GSD and the MaStar libraries appear when we separate the galaxies into star formation status bins (see above) and when we adopt the gradient slope to measure the metallicity at different radial distances.

In CF21 the separation between ChEHs narrowed from the center to the outskirts, such that (i) the different galaxy mass bins present a smaller difference in metallicity and (ii) showing that low mass galaxies always had a positive metallicity gradient as opposed to high mass ones. Figs. \ref{fig:gsd_zh_pos}, \ref{fig:zh_pos} show the same result, the former for the GSD library and the latter for the MaStar library. The ChEHs measured at the outskirts are practically identical in the outskirts for GSD while for MaStar there is still some segregation with mass. The ChEHs for the CALIFA sample at the outskirts show a segregation too, so this is not only an effect of the stellar library. Indeed, the CALIFA sample covers a larger average galactocentric radius compared to MaNGA (as a result of instrument FOV and sample selection), which combined with the narrower range in metallicity values of the GSD library can explain why the ChEHs measured at the outskirts of the MaNGA sample with GSD are less reliable.

A key difference arises between the results for GSD and MaStar for the metallicity gradient. For the GSD library high stellar mass galaxies have negative (in-out) values of the metallicity gradient and low mass galaxies have positive (out-in) values instead, for all cosmic times. For the MaStar library, however, all galaxies currently have a negative metallicity gradient (shallower for low mass ones) but low mass galaxies used to have a positive one, signaling a transition.
\section{Conclusions}
\label{sec:conclusions}

In this work we present the chemical evolution history of galaxies in the MaNGA sample measured using the spectral synthesis technique. The methodology employed allows us to analyze the same set of galaxies throughout cosmic time which yields a more consistent evolution. We find that stellar mass segregates the ChEH of galaxies both in terms of the value of their metallicity and of how quickly they become enriched, with the more massive galaxies having higher metallicities and a quicker evolution.

This is in line with previous works, but the dependence on stellar mass becomes less important after we separate the sample into morphology bins. Whereas the earlier type galaxies show a similar dependence on stellar mass the later morphological types become progressively less dependent on mass. The signature for this effect is the global flattening of the MZR which inverts for Sd galaxies. The delay in enrichment experienced by low mass galaxies compared to the high mass ones is also affected in a similar way.

We also compare the chemical evolution of galaxies depending on their SFS. For the ChEH we find results consistent with the correlation between morphology and SFS but the MZR shows a unique feature which only appears to depend on SFS: The high and low mass flattenings of the MZR (both of which appear for the full sample) are associated with either RG or SFG. The high mass flattening appears in RG but not in SFG and vice-versa for the low-mass flattening.

Another key result is found when comparing the ChEHs at the center and at the outskirts of the galaxies: We detect an inversion of the metallicity gradient such that galaxies below $10^{10}$ M$_\sun$ switch from an outside-in growth to an inside-out one. Higher mass galaxies either maintained inside-out growth throughout their lifetimes or we are unable to observe the switch with our method, due to the lesser reliability of older populations in the spectral synthesis technique.

As a result of the averaging technique employed we can separate the variance in the ChEHs of the galaxy bins we employ into "Scale" and "Shape". The former is the variance related to the difference in the current value of the metallicity, while the latter is the variance related to how different the shapes of the ChEHs are. We find that in general galaxies are more diverse in terms of the value of their metallicity rather than the rate of enrichment and that the variance grows for lower masses and later morphological types.

\acknowledgments
We are grateful for the support of a CONACYT grant CB-285080 and FC-2016-01-1916, and funding from the PAPIIT-DGAPA-IN100519, PAPIIT-DGAPA-IN103820, PAPIIT-DGAPA-IG100622, PAPIIT-DGAPA-IN112620, and PAPIIT-DGAPA-IA100420 (UNAM) projects.
GAB gratefully acknowledges support by the ANID BASAL project FB210003.

Funding for the Sloan Digital Sky Survey IV has been provided by the Alfred P. Sloan Foundation, the U.S. Department of Energy Office of Science, and the Participating Institutions. SDSS acknowledges support and resources from the Center for High-Performance Computing at the University of Utah. The SDSS web site is \url{www.sdss.org}.

SDSS is managed by the Astrophysical Research Consortium for the Participating Institutions of the SDSS Collaboration including the Brazilian Participation Group, the Carnegie Institution for Science, Carnegie Mellon University, Center for Astrophysics $\vert$ Harvard \& Smithsonian (CfA), the Chilean Participation Group, the French Participation Group, Instituto de Astrof\'{i}sica de Canarias, The Johns Hopkins University, Kavli Institute for the Physics and Mathematics of the Universe (IPMU) / University of Tokyo, the Korean Participation Group, Lawrence Berkeley National Laboratory, Leibniz Institut f\"{u}r Astrophysik Potsdam (AIP), Max-Planck-Institut f\"{u}r Astronomie (MPIA Heidelberg), Max-Planck-Institut f\"{u}r Astrophysik (MPA Garching), Max-Planck-Institut f\"{u}r Extraterrestrische Physik (MPE), National Astronomical Observatories of China, New Mexico State University, New York University, University of Notre Dame, Observat\'{o}rio Nacional / MCTI, The Ohio State University, Pennsylvania State University, Shanghai Astronomical Observatory, United Kingdom Participation Group, Universidad Nacional Aut\'{o}noma de M\'{e}xico, University of Arizona, University of Colorado Boulder, University of Oxford, University of Portsmouth, University of Utah, University of Virginia, University of Washington, University of Wisconsin, Vanderbilt University, and Yale University.

\vspace{5mm}
\facilities{HST(STIS), Swift(XRT and UVOT), AAVSO, CTIO:1.3m,
CTIO:1.5m,CXO}

\software{astropy \citep{AstropyCollaboration2013},  
        %   Cloudy \citep{2013RMxAA..49..137F}, 
        %   SExtractor \citep{1996A&AS..117..393B}
          }

%% Appendix material should be preceded with a single \appendix command.
%% There should be a \section command for each appendix. Mark appendix
%% subsections with the same markup you use in the main body of the paper.

%% Each Appendix (indicated with \section) will be lettered A, B, C, etc.
%% The equation counter will reset when it encounters the \appendix
%% command and will number appendix equations (A1), (A2), etc. The
%% Figure and Table counter will not reset.

\appendix

\section{Stellar library sampling and its impact on the results}
\label{sec:appendix-libraries}
There are four stellar libraries based on MaStar spectra that we considered for use in this study (LOG, LIN, MIX, sLOG), as well as the GSD \citep[GSD156][]{CidFernandes2013} stellar library. We use the latter in order to have a direct comparison between the results for the CALIFA sample presented in CF21 and those of this study (see Appendix B).

The results presented here are the first that use these stellar templates derived using the MaStar stellar library, and thus we performed tests regarding the suitable sampling of the parameter space, mainly that of the age of the populations. We present here a comparison and discussion of how the sampling in age affects the results.

The stellar template library we determined to be optimal is sLOG and thus all results presented in the main body of the article were produced using this library. All four MaStar libraries comprise 7 metallicities, ranging from Z = 0.0001 to Z = 0.04, differing only in the age sampling. The LOG and LIN libraries sample the ages in logarithmic and linear way, respectively. By construction the first one samples the recent ages in a more refined way, while the second one samples the time in an homogeneous way. The MIX library samples the ages below 1 Gyr using the LOG distribution, and above that age it uses the LIN one. Finally, the sLOG library adopts an intermediate procedure, sampling the age in multiplicative steps, in which the step is longer at larger ages.

The GSD stellar template library comprises 4 metallicities, ranging from Z = 0.004 to Z = 0.03. Its age sampling is most similar to that of LOG and sLOG among the MaStar SSP libraries.

 \begin{figure*}
	\includegraphics[width=\linewidth]{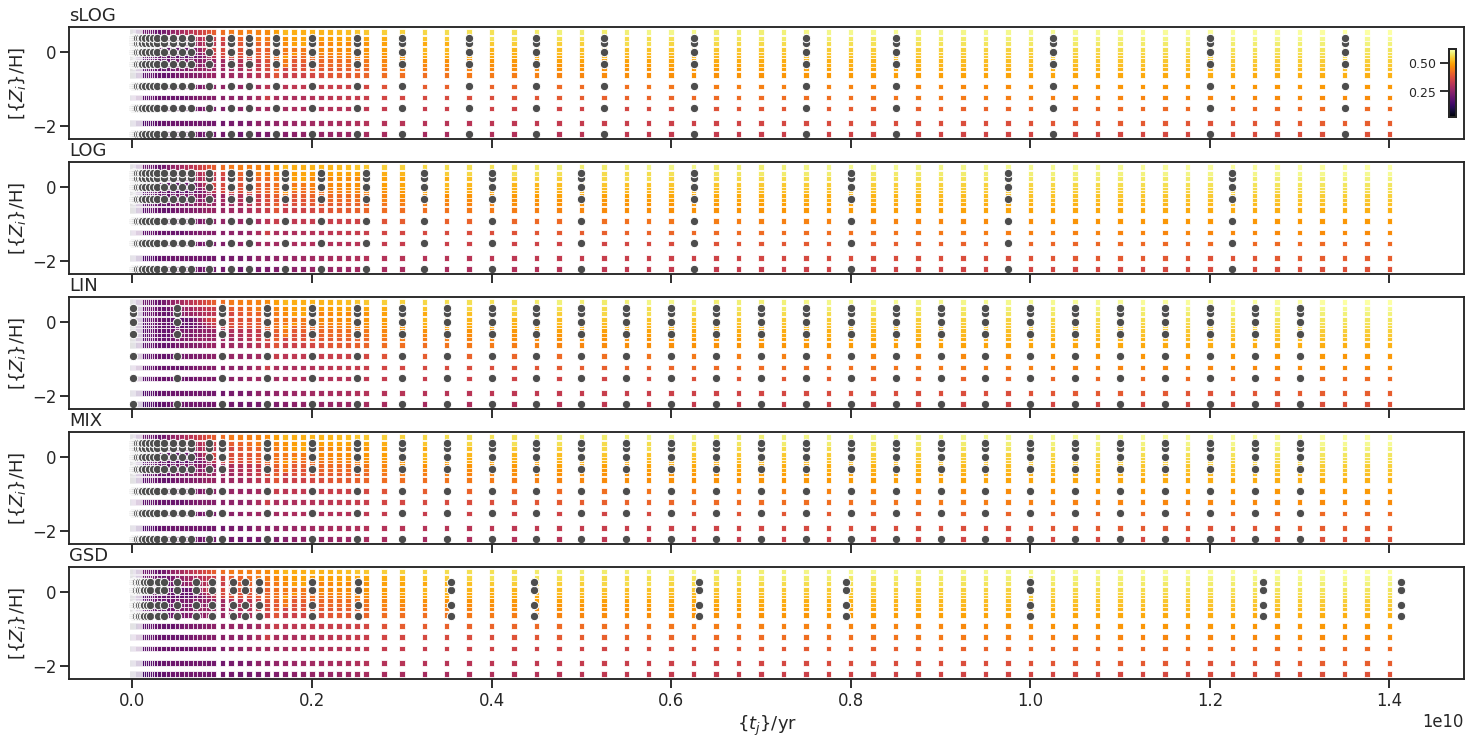}
     \caption{Representation of the sampling of the five stellar population templates considered in this paper, from top to bottom: sLOG, LOG, LIN, MIX and GSD (black dots), in the metallicity vs age plane. The full templates for the MaStar stellar library without age and metallicity selection are also shown as colored squares. The color represents a function of the typical distance between them in terms of similarity of their spectra. We use this distance as a proxy for the degeneracy between these templates.
     }
     \label{fig:templates}
 \end{figure*}
 
In Fig. \ref{fig:templates} we show a comparison between the five different libraries considered in this article and how their sampling differs. The age-metallicity values of the templates selected are represented in black dots over the full templates, which are represented as colored squares. The color represents a sort of distance between the templates in terms of how similar their spectra are, calculated using the chi-square between a template spectra and those of its closest neighbors. The GSD templates are the only ones where the black dots representing the selection do not match the positions of the full templates, which is natural since it is the only SSP library which was not selected from the models of the MaStar stellar library.

We have measured the ChEH and MZR in the same manner as in the main body of the article using these different stellar libraries for the sake of completeness and as an exercise to assess the impact that the sampling has on our data.

 \begin{figure*}
 	\includegraphics[width=\linewidth]{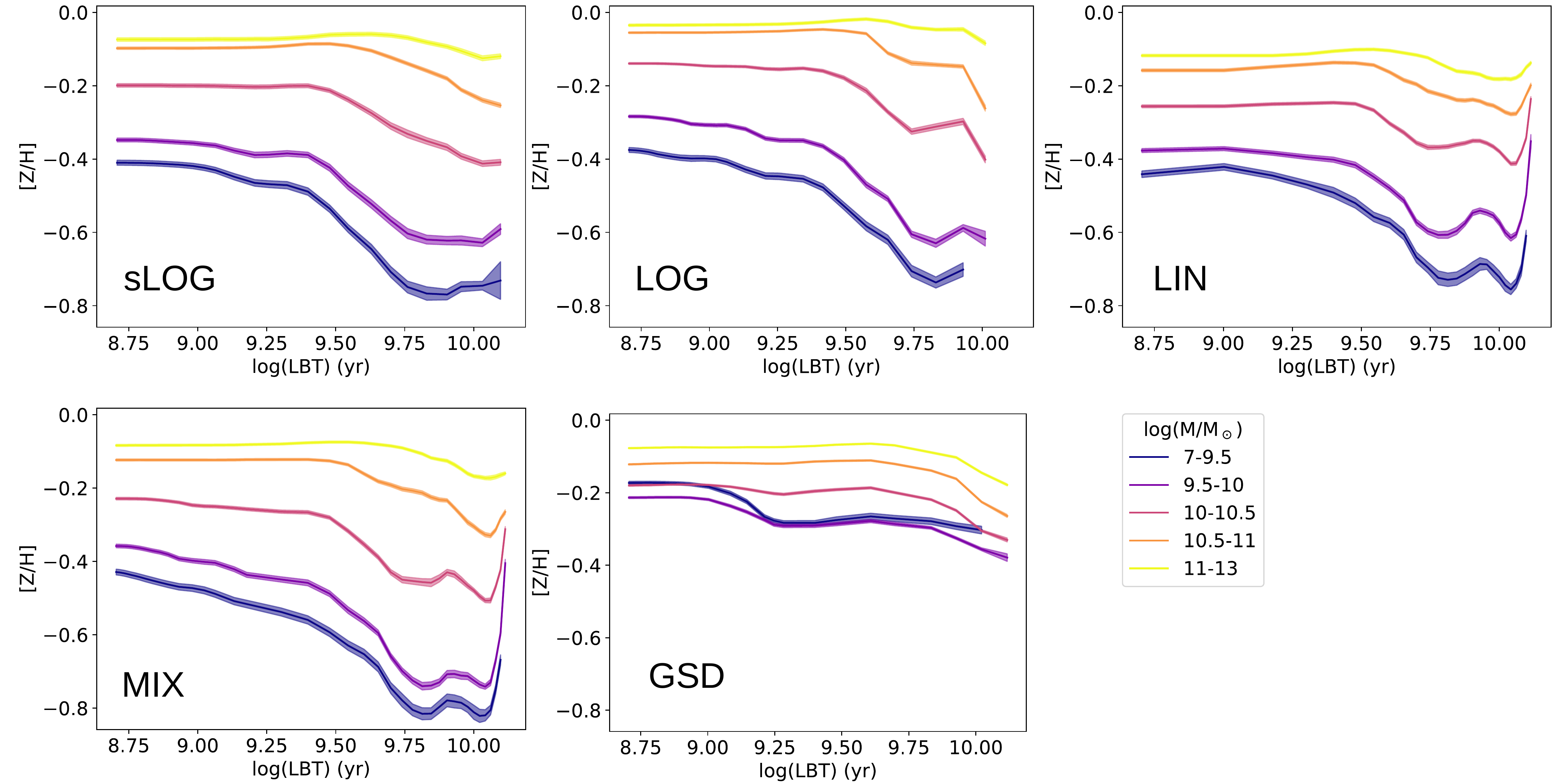}
     \caption{Comparison of the ChEH which results from using different stellar libraries. sLOG is the main library used in this work, while LOG, LIN and MIX are libraries produced with the same spectra but different samplings in age. Note that the GSD templates were not produced using the MaStar stellar library.
    }
     \label{fig:zh_comp}
 \end{figure*}

 \begin{figure*}
 	\includegraphics[width=\linewidth]{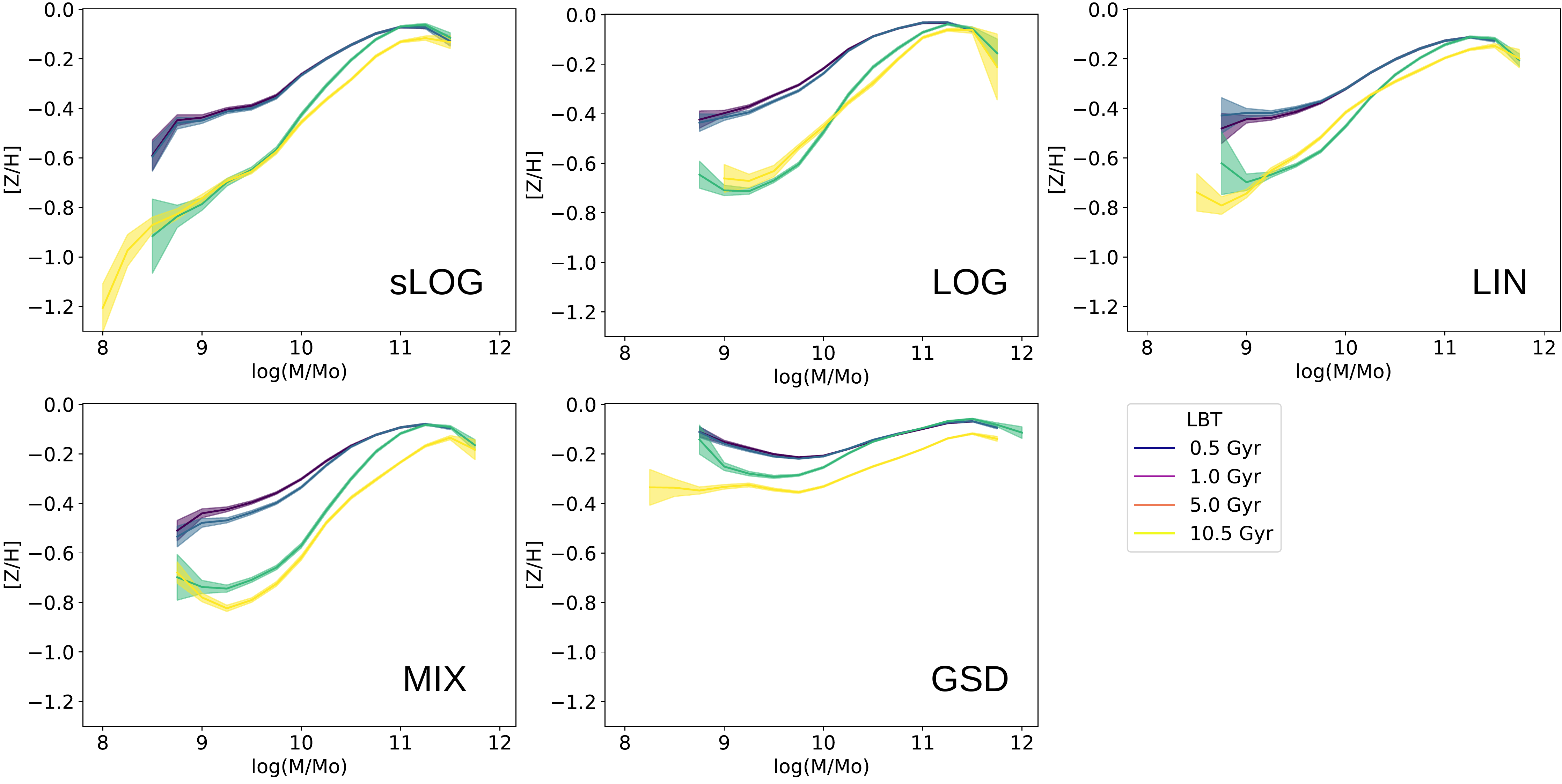}
     \caption{Comparison of the MZR evolution which results from using different stellar libraries. sLOG is the main library used in this work, while LOG, LIN and MIX are libraries produced with the same spectra but different samplings in age. Note that the GSD templates were not produced using the MaStar stellar library.}
     \label{fig:mzr_comp}
 \end{figure*}

In Figures \ref{fig:zh_comp} and \ref{fig:mzr_comp} we compare the ChEHs and MZR which result from using different stellar template libraries.

The results are generally compatible between the libraries that are based on the same stellar spectra and that cover the same metallicity range (sLOG, LOG, LIN, MIX), though there are some differences in the absolute values of the metallicity, with LOG showing the highest values and LIN the lowest. The GSD library has very different values for the metallicity for lower mass bins, but this is a result of the narrower range in metallicity values that the library has compared to the others. However, the same general results regarding the shape of the ChEH can be seen, there is no "saturation" in terms of the metallicity as one might expect. For example, the three lowest mass bins' metallicities in the MaStar libraries results are below the minimum values of the metallicity measured for the GSD library, but their ChEhs in the GSD library do not collapse to the same metallicity.

The LIN and MIX libraries show an odd feature in the ChEHs for the earliest LBT, an upturn of the metallicity towards earlier LBT which sharply rises. The degree to which the metallicity would then have dropped from the earliest times, when galaxies were still in their growth phase, suggests this is a spurious result. This is further reinforced by the fact that this feature disappears completely if the age sampling is changed. Both MIX and LIN have similar age samplings for the oldest populations.
It is very likely that this spike in metallicity is the result of the degeneracies intrinsic to the method. Older populations need to have a wider sampling because their spectra change less between over time. We should recall that both libraries sample ages older than 1 Gyr in a linear way. This is strongly not recommended, since the differences between the spectra of old stellar populations does not allow us to make a clean distinction between them \citep[see][]{Bruzual2003,CidFernandes2014,Conroy2013}.

The sLOG library was created to avoid this issue while adding some more data points for the older populations compared to LOG. We consider it to represent the best balance, showing consistent results with LOG but better characterized for the further values of the LBT, as can be observed in Figure \ref{fig:zh_comp}.

In regards to the MZR, the results are mostly similar except for those regarding the low mass end of the MZR. All libraries except for sLOG have an inversion of the MZR at the very lowest masses for LBT older than 1 Gyr. We have no explanation for this but it should be noted that this happens for mass values that have a significantly lower number of galaxies and therefore a higher error in the determination of the MZR.

\section{Results using GSD library for direct comparison between CALIFA (CF21) and MaNGA results} \label{sec:CALIFA}

 \begin{figure}
 	\includegraphics[width=\columnwidth]{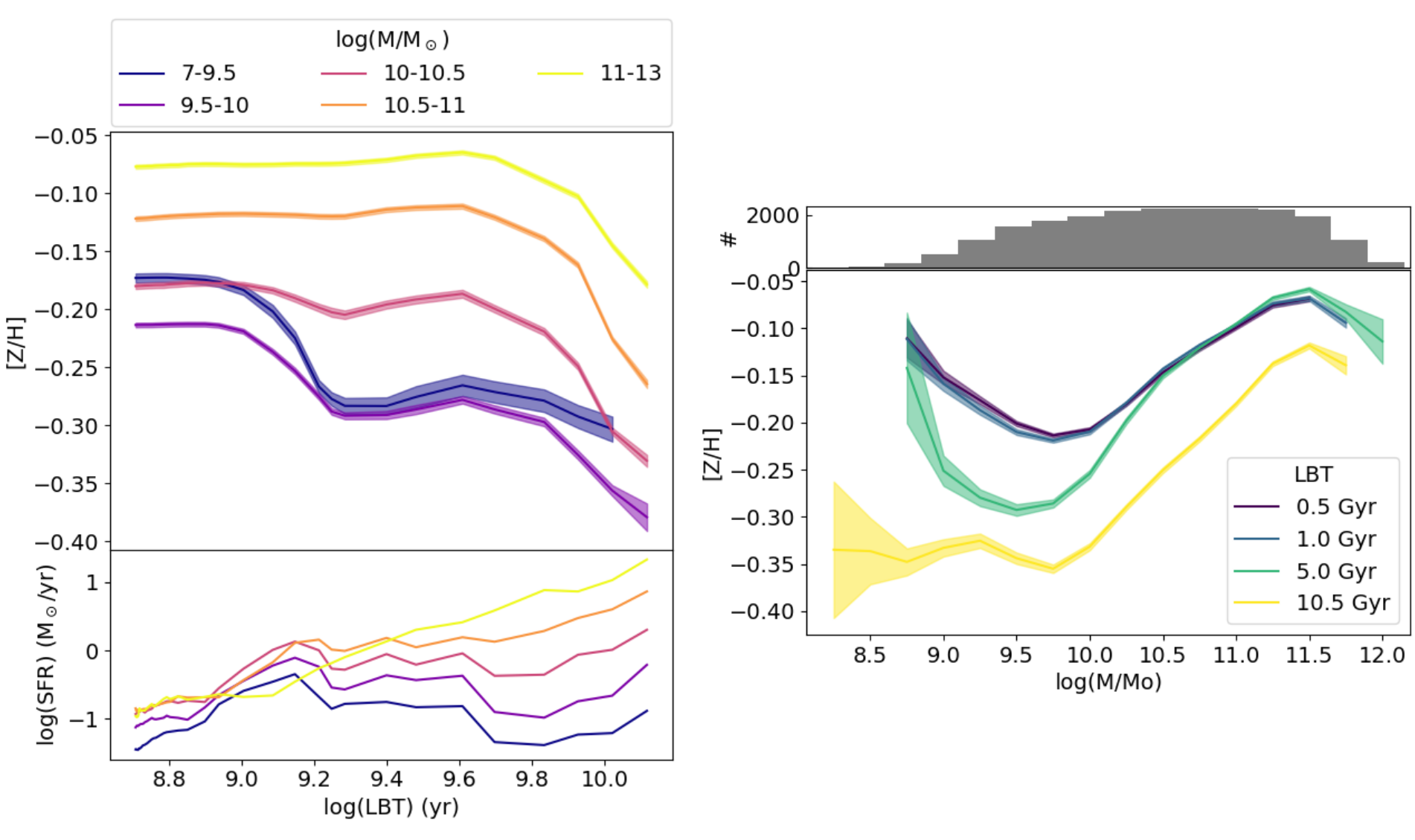}
     \caption{In the left, the evolution of the chemical enrichment along cosmic time for all galaxies in our sample for the GSD stellar library (bottom panel shows the SFH). In the right, the evolution of the MZR along cosmic times for the GSD stellar library (top panel shows the distribution of the sample in currently observed stellar mass).}
     \label{fig:gsd_zh_all}
 \end{figure}

 \begin{figure*}
 	\includegraphics[width=\linewidth]{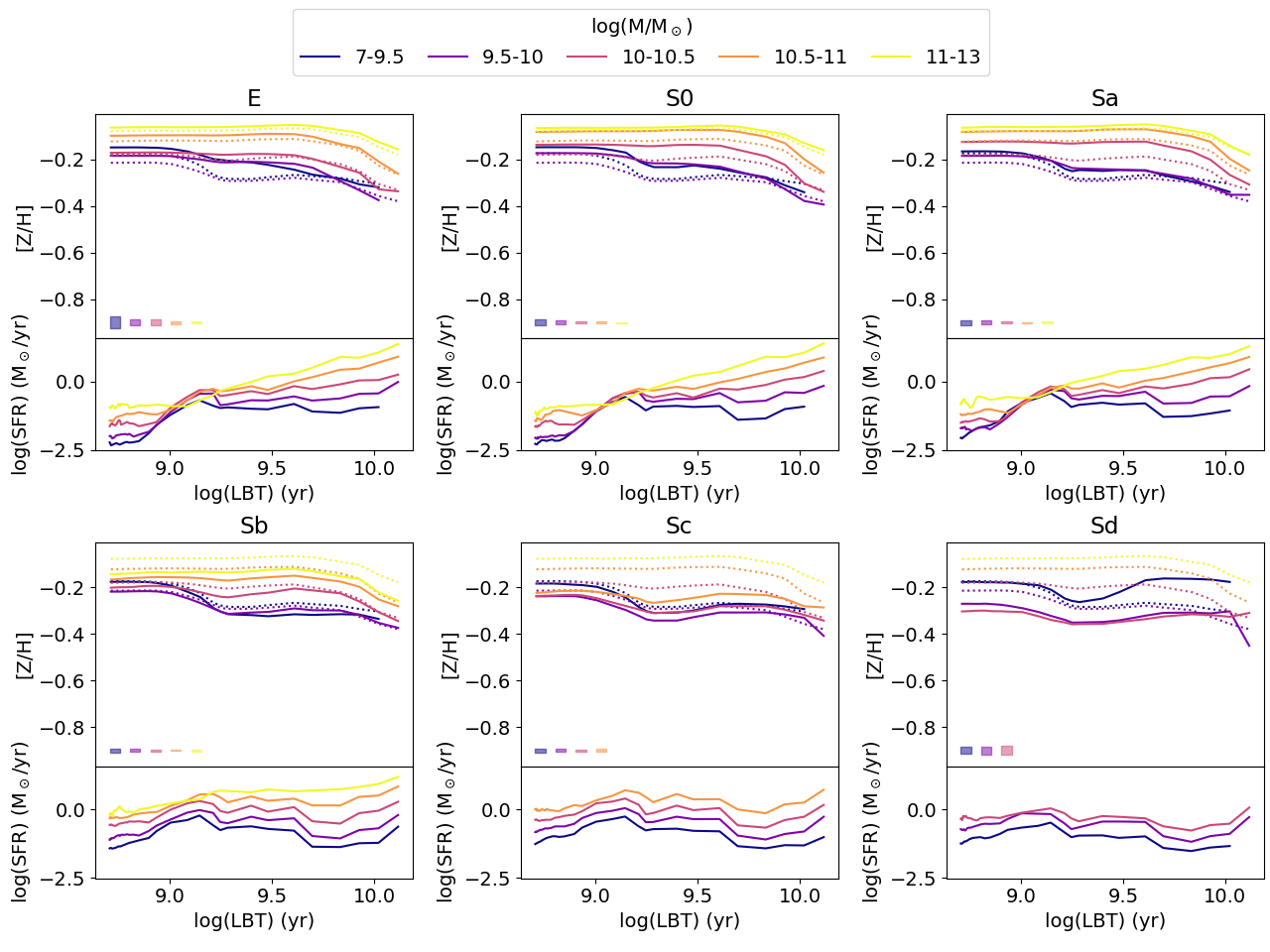}
     \caption{Evolution of the chemical enrichment for all galaxies in our sample for the GSD stellar library. On the bottom panel we show the SFH.}
     \label{fig:gsd_zh_morph}
 \end{figure*}

  \begin{figure*}
 	\includegraphics[width=\linewidth]{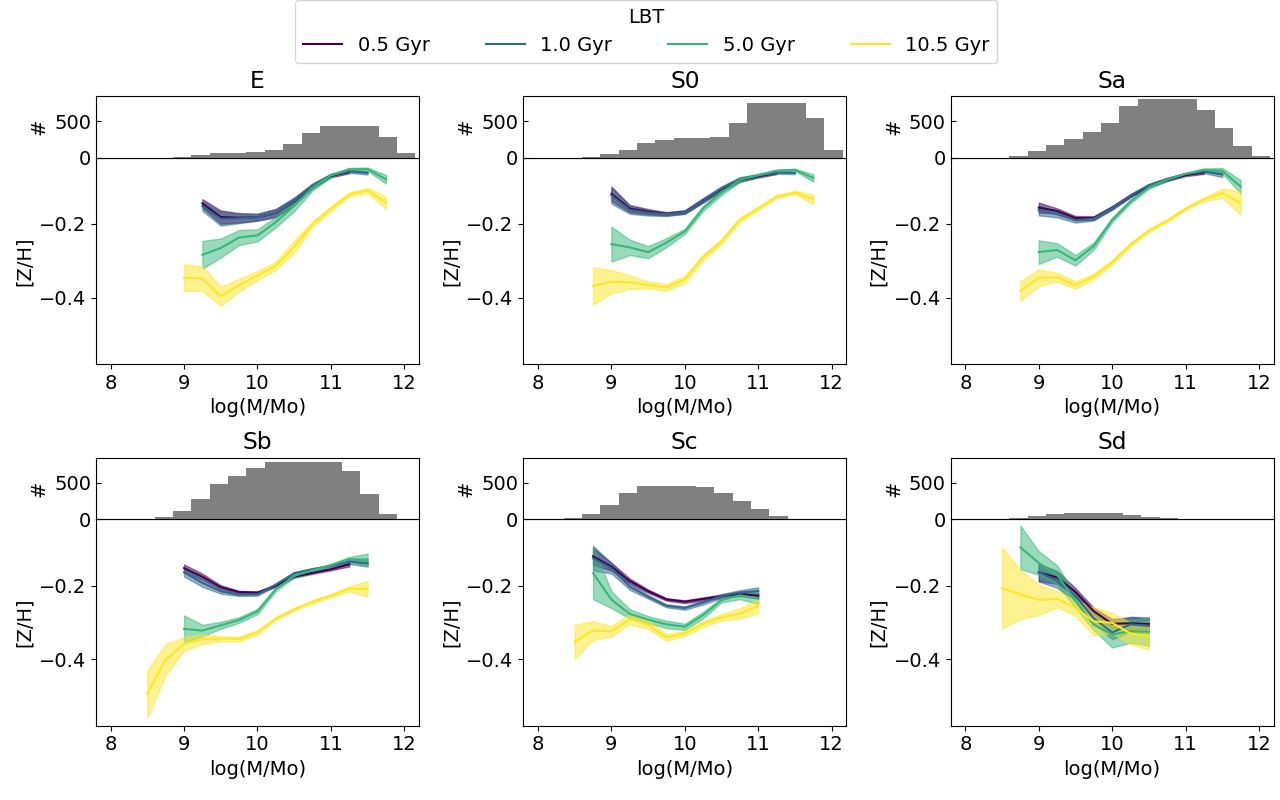}
     \caption{Evolution of the MZR for all galaxies in our sample for the GSD stellar library. On the top panel we show the distribution of the sample in currenty observed mass.}
     \label{fig:gsd_mzr_morph}
 \end{figure*}

 \begin{figure*}
 	\includegraphics[width=\linewidth]{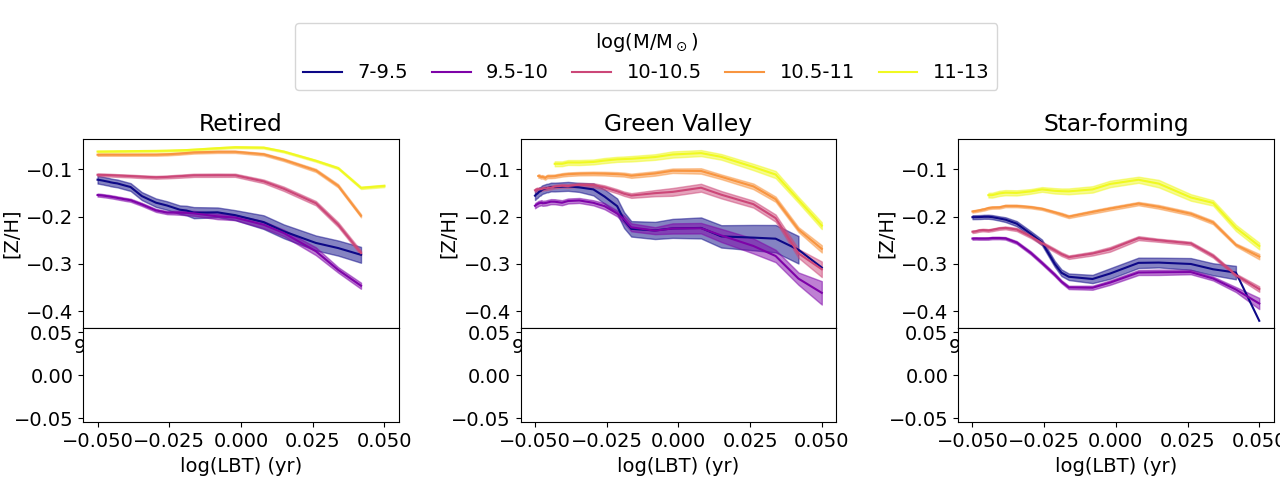}
     \caption{Evolution of the chemical enrichment for all galaxies in our sample for the GSD stellar library. On the bottom panel we show the SFH.}
     \label{fig:gsd_zh_sf}
 \end{figure*}
 
  \begin{figure*}
 	\includegraphics[width=\linewidth]{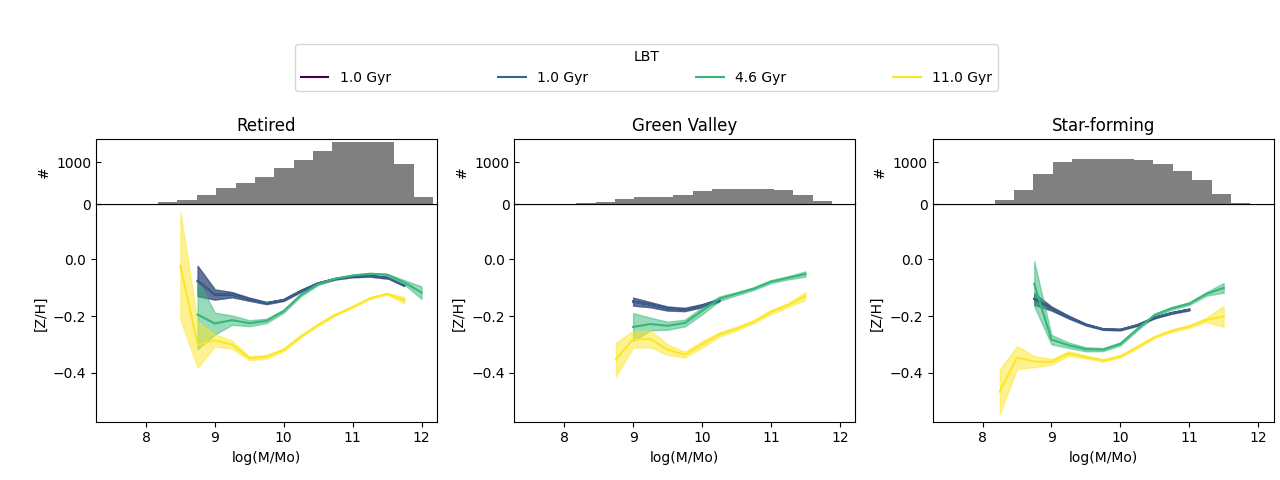}
     \caption{Evolution of the MZR for all galaxies in our sample for the GSD stellar library. On the top panel we show the distribution of the sample in currenty observed mass.}
     \label{fig:gsd_mzr_sf}
 \end{figure*}

 \begin{figure*}
 	\includegraphics[width=\linewidth]{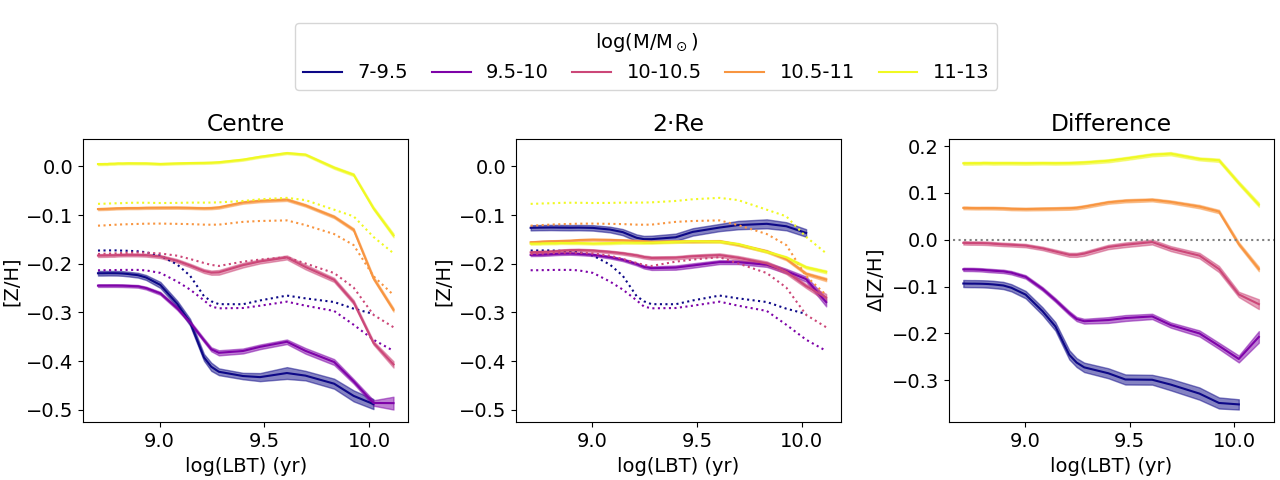}
     \caption{Evolution of the chemical enrichment for all galaxies in our sample for the GSD stellar library. On the bottom panel we show the SFH.}
     \label{fig:gsd_zh_pos}
 \end{figure*}

 \begin{figure*}
 	\includegraphics[width=\linewidth]{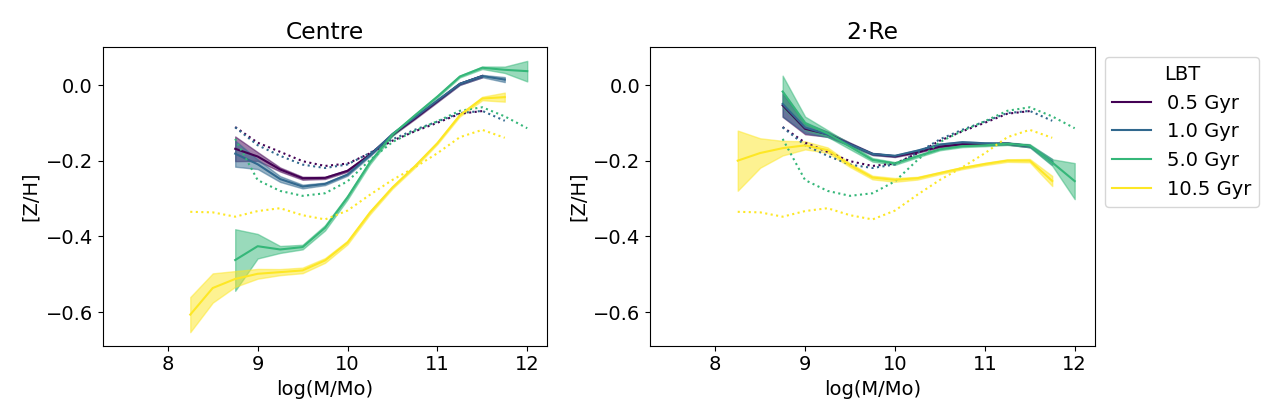}
     \caption{Evolution of the MZR for all galaxies in our sample for the GSD stellar library. On the top panel we show the distribution of the sample in currenty observed mass.}
     \label{fig:gsd_mzr_pos}
 \end{figure*}

 \begin{figure*}
 	\includegraphics[width=\linewidth]{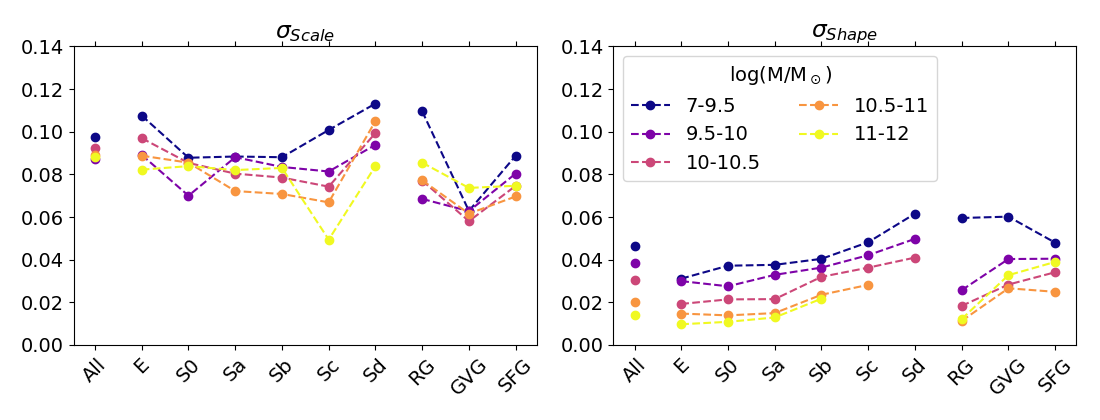}
     \caption{Comparison of the standard deviations of the ChEH for the different groups considered in this article using the GSD stellar library, separated into the standard deviation due to the absolute value differences in metallicity (Scale) and that due to a difference in the shape of the ChEH (Shape).}
     \label{fig:gsd_variance}
 \end{figure*}

\bibliography{export-bibtex}{}
\bibliographystyle{aasjournal}

%% This command is needed to show the entire author+affiliation list when
%% the collaboration and author truncation commands are used.  It has to
%% go at the end of the manuscript.
%\allauthors

%% Include this line if you are using the \added, \replaced, \deleted
%% commands to see a summary list of all changes at the end of the article.
%\listofchanges

\end{document}